\documentstyle[aps,eqsecnum,pra,epsfig]{revtex}

\newcommand{\krn}{\kern -2pt}
\newcommand{\ra}{{\Big \rangle}}
\newcommand{\la}{{\Big \langle}}

\newcommand{\edm}{\mbox{$H_{_{\rm PTV}}$}}
\newcommand{\he}{\mbox{$V_{\rm es}$}}

\newcommand{\m}{\Big |}
\def\r{\rangle}
\def\l{\langle}

\begin{document}

\title{Coupled Electron Pair Approximation Calculation of the 
       Electric Dipole Moment of Atomic Yb}
\author{ Angom Dilip Singh}
\address{Physical Research Laboratory,           \\
         Navarangpura, Ahmedabad--380 009.
         }
\author{ Bhanu Pratap Das}
\address{Non--Accelerator Particle Physics Group, \\
         Indian Institute of Astrophysics,        \\
         Sarjapur Road, Koramangala,              \\
         Bangalore-34 
         }
\author{ Debashis Mukherjee}
\address{Department of Physical Chemistry,\\
         Indian Association for Cultivation of Science , \\
         Calcutta-32
         }
        
\maketitle

\begin{abstract}
  The existence of a finite electric dipole moment (EDM) $d_a$ of the 
closed-shell atom Yb implies parity and time reversal violations involving the 
nuclear sector. An important effect which can contribute to the Yb EDM is the 
tensor-pseudotensor electron-nucleus interaction characterized by the coupling 
constant $C_T$. Within the Standard Model (SM) of particle physics $C_T=0$, as 
this form of interaction is not allowed. If a finite $d_a$ of Yb is observed 
in experiments, then an estimate of $C_T$ can be obtained by combining with the
theoretical calculations. A non-zero $C_T$ implies physics beyond the standard 
model. In this paper we present the result of our {\em ab initio} calculation of
the EDM  Yb using different many-body methods.
\end{abstract}


\section{Introduction}

  Discrete symmetry violations in atoms are important phenomena to probe for 
physics beyond the Standard Model(SM) of the particle physics. The electric 
dipole moment(EDM) of an atom, which is a signature of the simultaneous parity
and time-reversal (P-T) symmetry violations is one such. There are several
possible sources of P-T violation effects within an atom, the tensor
pseudo-tensor electron nucleus interaction is an example which is semi-leptonic
in nature. The atomic Yb, which is a closed-shell atom having a $Z=70$ is a
very good candidate for atomic EDM experiments to probe for the nuclear
sector effects. An additional advantage of using a rare Earth atom like Yb is 
the closely spaced energy levels. The  experimental results when compared with 
the theoretical results can yield signatures of physics beyond the SM. 

  Among the closed-shell atoms, the atom EDM of Xe\cite{vold} and 
Hg\cite{romalis1} have been measured. Though the Hg experiment has set the 
record of being the most sensitive spectroscopy ever done, the result obtained 
is null and sets an upper bound to atomic Hg EDM as $ 2.1\times 10^{-28}ecm$. 
Still, it has set bounds on the parameters in particle physics 
models\cite{romalis1}. Improving the accuracy further is an important
challenge in atomic EDM experiments. Atomic Yb offer a possibility of achieving
this by using the techniques of laser cooling and trapping. It is also
desirable that the EDM measurements be done in other closed-shell atoms to 
verify the physical effects observed. 

  Atomic Yb also has the advantage of being relatively simple in level
structure among the rare Earth elements. It has been studied theoretically
using a variety of atomic many-body methods and there are ongoing 
experimental studies. The excitation energies has been studied using 
relativistic coupled-cluster method\cite{kaldor}; the hyperfine structure 
constants and electric-dipole transition properties have been studied using 
multireference relativistic many-body perturbation 
theory\cite{porsev1,porsev2}; and the oscillator strength of some of the 
important transitions were investigated using multi-configuration relativistic 
Hartree-Fock\cite{migdalek}. The lifetime of some of the crucial levels have 
been measured experimentally\cite{budker} and atomic Yb has been laser cooled 
and trapped\cite{watanabe,honda}. It is the candidate for the atomic EDM 
experiments using the methods of laser cooling and trapping of atoms which 
are in progress\cite{takahashi,romalis2}.

  The P-T violating effects which can be probed using atomic Yb are the tensor
pseudotensor (T-PT) electron-nucleus interaction\cite{barr} and the 
Schiff moment\cite{flambaum}. An observation of a finite EDM of atomic Yb 
could mean nonzero $C_T$, which is a signature of physics beyond the SM as 
$C_T$ is zero within SM. Hence the theories which allow T-PT electron-nucleus 
interactions would play a role in understanding nature. The parameter $C_T$
is extracted from the experimental result of atomic EDM by combining with the
theoretical calculations. Accurate atomic theory calculations are needed to
obtain precise $C_T$.The atomic EDM can also put bounds on the parameters of 
alternative models in particle physics. The Schiff moment which arises due to 
either nucleon EDM or P-T violating interactions between the nucleons can also 
contribute to the Yb EDM and can be used to extract the nuclear P-T violation 
parameters. We have calculated the contribution to atomic Yb EDM from both 
these effects but in this paper we will present the result of our calculations 
for T-PT contribution alone, in a later paper we will present the result of 
our Schiff moment Computation.

  The calculation of Yb EDM within a limited set of configuration state
functions (CSFs) using diagonalization and Bloch equation based many-body
perturbation theory ( MBPT) methods was reported in an earlier 
paper\cite{angom}. For convenient of reference this paper is called as paper-I 
hereafter. As a followup, in this paper we present results of 
calculations using much larger CSF space. In addition, we discuss the 
drawbacks common to the diagonalization and Bloch equation based 
perturbation methods, improvements are discussed and results of using these 
methods are presented.  In this paper we present the method of the calculation.


\section{Bloch Equation Based Many-Body Perturbation Theory}

  This section is a brief overview of the interaction used in the 
calculation and Bloch equation based MBPT. A detailed description of which are 
given in the paper-I. The effective T-PT electron-nucleus interaction 
Hamiltonian obtained by treating the nucleus nonrelativistically is 
\begin{equation}
  H_{_{\rm PTV}}= i 2\sqrt{2}\bigg ( C_TG_F\bigg )
  \bigg ( \vec{I}\!\cdot\! \beta\vec{\alpha}\bigg ) \rho_N(r),
\end{equation}
where $C_T$ is the T-PT electron-nucleus coupling constant, $G_F$ is the Fermi
coupling constant, $\vec{I}$ is the nuclear spin, $\beta$ and $\alpha$ are the
Dirac matrices and $\rho_N(r)$ is the nuclear density. This interaction 
Hamiltonian is effective only within the nuclear region, where $\rho_N(r)$ is 
non-zero. And the dependence on $\vec{I}$ implies that it is observable only in 
odd isotopes of Yb which has nonzero $\vec{I}$. We use Fermi nuclear density 
model for our calculations. 

  The Bloch equation of an atom with unperturbed Hamiltonian $H_0$ and the 
residual coulomb interaction $V_{\rm es}$ as the perturbation is 
\begin{equation}
  \left [ \Omega_{\rm es}, H_0\right ] = V_{\rm es}\Omega_{\rm es}
                                  -\chi_{\rm es}PV_{\rm es}\Omega_{\rm es}.
$$
\end{equation}
Where $\Omega_{\rm es}$ is the wave-operator, $\chi_{\rm es}= \Omega_{\rm es}
-1 $ is the correlation operator, $P$ and $Q$ are the projection operators of
the model and complementary space. The first and second terms on the right 
hand side are the principal and renormalization terms respectively. Using
Epstein-Nesbet(E-N) partitioning $H_0$ and $V_{\rm es}$ can be defined in terms
of a set of CSFs as 
\begin{eqnarray}
  H_0 &= &\sum_i \la\Phi_i\m H_{\rm atom}\m\Phi_i\ra\m\Phi_i\ra\la\Phi_i\m
        +\sum_i \la\overline{\Phi}_i\m H_{\rm atom}\m\overline{\Phi}_i\ra\m
        \overline{\Phi}_i\ra\la\overline{\Phi}_i\m , \nonumber  \\
  V_{\rm es} &= &\sum_{ij} \la\Phi_i\m H_{\rm atom}\m\Phi_j\ra\m\Phi_i\ra
        \la\Phi_j\m +\sum_{ij} \la\overline{\Phi}_i\m H_{\rm atom}\m
        \overline{\Phi}_j\ra\m \overline{\Phi}_i\ra\la\overline{\Phi}_j\m ,
                      \nonumber 
\end{eqnarray}
where $H_{\rm atom}$ is the Dirac-Coulomb atomic Hamiltonian, and  
$\{|\Phi_i\rangle\}$ and $\{|\overline{\Phi}_i\rangle\}$ are the even and odd 
parity CSF spaces respectively. The Dirac-Coulomb Hamiltonian of an atom with
$N$ electrons in atomic units ( $e=1$, $\hbar=1$, and $m_e=1$ ) is 
\begin{equation}
 H_{\rm atom} = \sum_{i=1}^N \bigg ( c\vec{\alpha}_i\cdot\vec{p}_i +
                (\beta_i-1)c^2 - V_{\rm nuc}(r_i)\bigg ) + \frac{1}{2}
                \sum_{i,j}^{N,N}\frac{1}{r_{ij}},
\end{equation}
where $\vec{\alpha}_i$ is the Dirac matrix, $\vec{p}_i$ is the momentum
of the electron and $V_{\rm nuc}$ is the nuclear potential. Introduce the P-T 
violating interaction Hamiltonian $H_{_{\rm PTV}}$ as perturbation and define 
the total perturbation Hamiltonian $V$ as 
\begin{equation}
  V = V_{\rm es} + H_{_{\rm PTV}},
\end{equation}
The interaction Hamiltonian $H_{_{\rm PTV}}$ can also be expressed in terms
of CSFs as in $H_0$ and $V_{\rm es}$. Treating $V$ as the perturbation
and defining $\Omega({\rm edm})$ as the total wave operator
\begin{equation}
  \Omega({\rm edm}) = \Omega_{\rm es} + \Omega_{\rm es,edm},
\end{equation}
where $\Omega_{\rm es}$ remains the same and $\Omega_{\rm es,edm}$ is the
wave-operator which has one order of $H_{_{\rm PTV}}$ and all possible orders
of $V_{\rm es}$ before and after $H_{_{\rm PTV}}$. The $H_{_{\rm PTV}}$ is 
treated to first order since it scales as $G_F$, which is very small. The 
expression of the atomic EDM $d_a$ is 
\begin{equation}
 d_a = \la\Phi_0\m \Omega^{\dagger}_{\rm es}\vec{D}\Omega_{\rm es,edm}\m\Phi_0
       \ra+ \la\Phi_0\m \Omega^{\dagger}_{\rm es,edm}\vec{D}
       \Omega_{\rm es}\m\Phi_0 \ra = \la\Phi_0\m \vec{D}_{\rm eff}\m\Phi_0\ra,
\end{equation}
where $\vec{D}_{\rm eff} = \Omega^{\dagger}_{\rm es}\vec{D}\Omega_{\rm es,edm}
+ \Omega^{\dagger}_{\rm es,edm}\vec{D} \Omega_{\rm es}$ is the effective 
atomic EDM operator. It is the dipole operator dressed with the all order 
residual coulomb interaction and one order of $H_{_{\rm PTV}}$ arranged in all 
possible sequence. The Bloch equation based MBPT has been used for the 
calculation with a large CSFs space as it is more efficient in terms of 
execution time unlike the direct matrix diagonalization approaches.


\section{Size Consistent Theory in Closed-Shell Systems}

  An atomic many-body theory is size consistent if the properties calculated 
using it scales linearly as the number of the electrons. The diagonalization and 
the Bloch equation based methods are size consistent within a complete 
CSF space. But it is size inconsistent if the configuration space considered is 
incomplete. A consequence of incomplete cancellation of the unlinked terms, 
which scales nonlinearly to the number of electrons. The cancellation is 
complete when the CSF space is complete. To study atomic EDM which has 
important implications and small in magnitude it is preferable to use 
atomic-many body theory which is size-consistent even with an incomplete CSF
space. It is also difficult to satisfy the condition of completeness for heavy 
atoms, which are important candidates for the atomic EDM experiments as the 
number of possible CSFs runs into millions for a moderate size orbital space.


\subsection{Size Consistency with Linked Diagram Theorem}
\begin{figure}[h]
  \centerline{
  \epsfig{height=1in,width=3in,figure=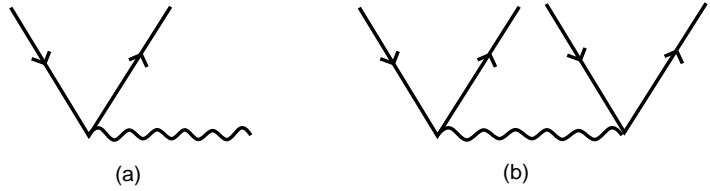}}
  \caption{ Diagrams for the wave-operators 
            (a)$\Omega_{\rm es}(1)$ and (b) $\Omega_{\rm es}(2)$. }
  \label{fig1}
\end{figure}

\begin{figure}[h]
  \epsfig{height=3in,width=\textwidth,figure=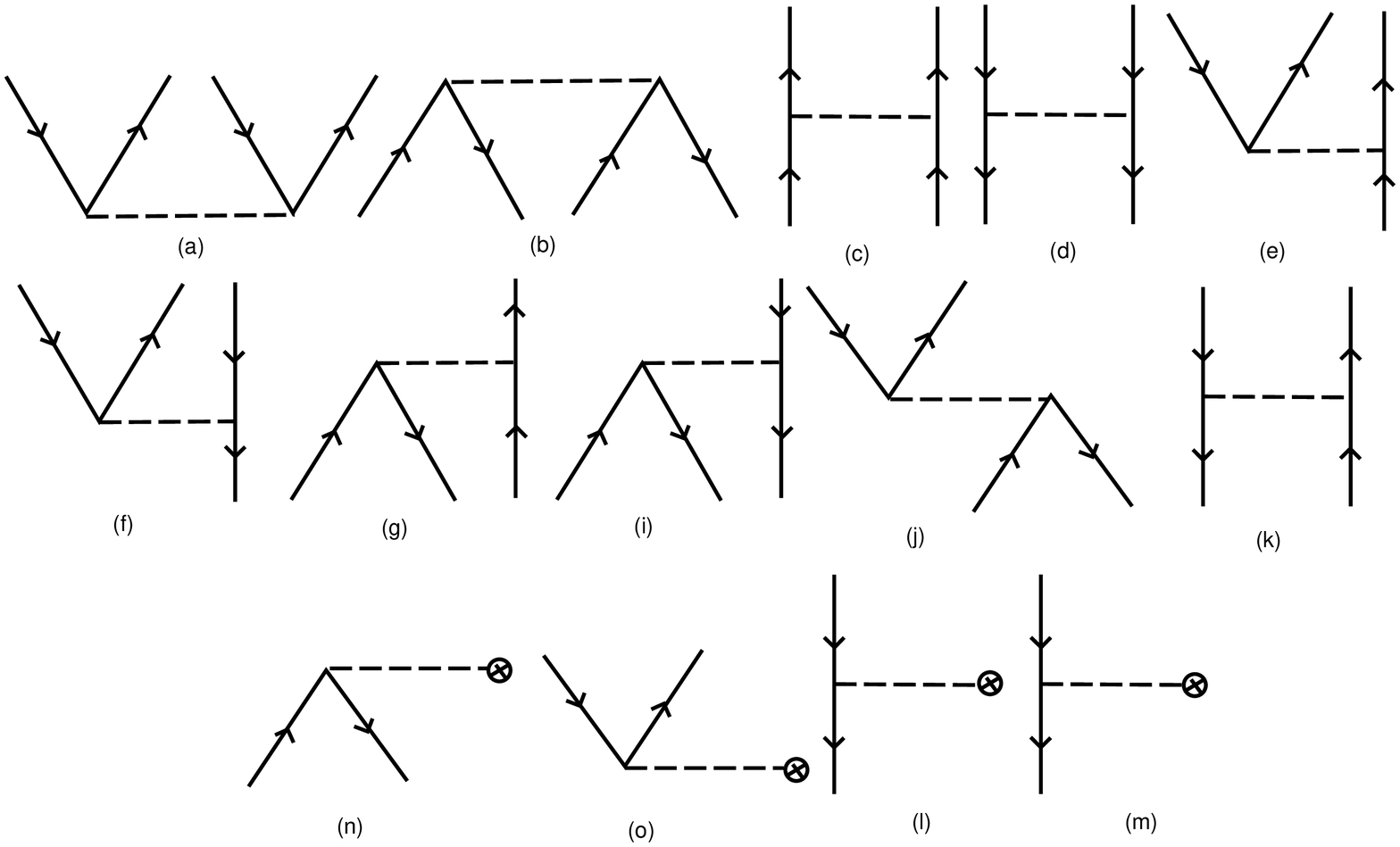}
  \caption{ The diagrams for the residual Coulomb interaction $\he $. }
  \label{fig2}
\end{figure}

  The wave-operator calculated using the Bloch equation within an incomplete 
CSF space is size-consistent if only the linked terms are 
retained\cite{goldstone}. The incomplete cancellation of the unlinked 
terms is then avoided
\begin{equation}
  \bigg [ \Omega_{\rm es}, H_0\bigg ]P = Q\bigg (\he \Omega_{\rm es}P - 
  \chi_{\rm es}P\he  \Omega_{\rm es}P\bigg  )_{\rm linked}
  = Q\bigg (\he \Omega_{\rm es}P - \chi_{\rm es}W \bigg  )_{ \rm linked},
 \label{eqn2.1}
\end{equation}
where $W = P\he  \Omega_{\rm es}P$. Redefine the wave-operator in terms of 
orders of excitation and consider only the single and double excitations. The 
wave-operator and correlation-operator are
\begin{equation}
  \Omega_{\rm es} = I + \Omega_{\rm es}(1) + \Omega_{\rm es}(2)
  = \sum_{m=0}^{2}\Omega_{\rm es}(m) \,\,\,\, \mbox{\rm and } \,\,\,\,
  \chi_{\rm es} = \sum_{m=1}^{2} \Omega_{\rm es}(m) ,
\end{equation}
where $m$ is the order of excitation. The closed-shell single and double 
excitation wave-operator diagrams are shown in Fig\ref{fig1}. Similarly, the 
$\he$ diagrams are shown in Fig\ref{fig2}. Unlike the single particle approach,
the diagrams are used as representation of the physical effects and cannot be 
evaluated directly using the usual Goldstone rules as E-N partitioning is
used. The $W$ is the energy and hence a number for closed-shell sytems. 
From the definitions
\begin{equation}
  \bigg [ \Omega_{\rm es}(m), H_0\bigg ] = Q\bigg (\he  + \he \Omega_{\rm es}
  (1) + \he \Omega_{\rm es}(2) - \Omega_{\rm es}(m)W\bigg  )_{ m,{\rm linked}}.
\end{equation}
Let $|\Phi_0\rangle $ be the reference configuration and 
$\{|\Phi_{\alpha}\rangle \}$ be the configuration space spanned by singly and 
doubly excited configurations. The wave-operators can then be expressed as 
\begin{equation}
  \Omega_{\rm es}(1) = \sum_{ar} \m\Phi_a^r \ra \la \Phi_0 \m x_a^r
  \,\,\,\, \mbox{\rm and }  \,\,\,\,
  \Omega_{\rm es}(2) = \sum_{abrs} \m\Phi_{ab}^{rs} \ra \la \Phi_0 \m x_{ab}^{rs}.
\end{equation}
Where $x_a^r$ and $x_{ab}^{rs}$ are the excitation amplitudes. For 
closed-shell systems with a single reference, $W$ is just a number or closed 
diagrams. The term $\Omega_{\rm es}W$ is  therefore unlinked and  does not 
contribute to the linked Bloch equation. The one-body wave-operator equation is
\begin{equation}
  \bigg [ \Omega_{\rm es}(1), H_0 \bigg ]P\!\! =\!\! \sum_{ar}\Bigg [ 
  \la \Phi_{a}^{r}\m \he \m\Phi_0\ra + \!\sum_{a'r'} \la \Phi_{a}^{r}\m \he \m
  \Phi_{a'}^{r'}\ra x_{a'}^{r'}+\!\!\sum_{a'b'r's'}\!\!\la\Phi_{a}^{r}\m\he\m 
  \Phi_{a'b'}^{r's'}\ra x_{a'b'}^{r's'} \Bigg ]_{\rm linked}
  \!\!\!\!\!\!\!\!\!\!\!\!\!\! \m\Phi_{a}^{r}\ra \la \Phi_0 \m
 \label{eqn2.2}
\end{equation}
Similarly, the two-body wave-operator the equation is 
\begin{equation}
  \bigg [ \Omega_{\rm es}(2), H_0 \bigg ]P\!\!=\!\! \sum_{abrs}\Bigg [ 
  \la \Phi_{ab}^{rs}\m \he \m\Phi_0\ra + \!\sum_{a'r'} \la \Phi_{ab}^{rs}\m 
  \he\m \Phi_{a'}^{r'}\ra x_{a'}^{r'} +\!\! \sum_{a'b'r's'} \!\!\la 
  \Phi_{ab}^{rs}\m \he \m \Phi_{a'b'}^{r's'} \ra x_{a'b'}^{r's'}\Bigg ]_{\rm 
  linked}\!\!\!\!\!\!\!\!\!\!\!\!\!\!\m\Phi_{ab}^{rs}\ra \la \Phi_0 \m
 \label{eqn2.3}
\end{equation}
Introduce $H_{_{\rm PTV}}$ as perturbation and redefine the perturbation
Hamiltonian as $V = \he + \edm$. The corresponding wave-operators are
\begin{equation}
  \Omega(1) = \Omega_{\rm es}(1) + \Omega_{\rm es,edm}(1) 
         \,\,\,\, \mbox{\rm and } \,\,\,\,
  \Omega(2) = \Omega_{\rm es}(2) + \Omega_{\rm es,edm}(2) .
\end{equation}
Wave-operators $\Omega_{\rm es}(1) $ and $\Omega_{\rm es}(2)$ are same as 
before but $\Omega_{\rm es,edm}(1)$ and $\Omega_{\rm es,edm}(2)$ 
connect $\{|\Phi_i\rangle \}$ to 
$\{|\overline{\Phi}_i\rangle \}$, where $\{|\overline{\Phi}_i\rangle \}$ is the
configuration space opposite in parity to $|\Phi_0\rangle $.  Within the total 
configuration space  $\Omega_{\rm es,edm}(1)$ and $\Omega_{\rm es,edm}(2)$ can 
be represented as
\begin{equation}
  \Omega_{\rm es,edm}(1) = \sum_{ar}\m\overline{\Phi}_a^r\ra \la \Phi_0\m
  \overline{x}_a^r \,\,\,\, \mbox{\rm and} \,\,\,\,
  \Omega_{\rm es,edm}(2) = \sum_{abrs}\m\overline{\Phi}_{ab}^{rs}\ra \la 
  \Phi_0\m \overline{x}_{ab}^{rs} 
\end{equation}
The equations of $\Omega_{\rm es,edm}(1)$ is
\begin{eqnarray}
  \bigg [ \Omega_{\rm es,edm}(1),H_0 \bigg ]P &= &\!\!\!\sum_{ar}\Bigg [ \la 
  \overline{\Phi}_a^r\m\edm\m\Phi_0\ra + \sum_{a'r'}\la \overline{\Phi}_a^r\m
  \edm\m\Phi_{a'}^{r'}\ra x_{a'}^{r'} + \!\!\sum_{a'b'r's'}\!\la
  \overline{\Phi}_a^r\m\edm\m\Phi_{a'b'}^{r's'}\ra x_{a'b'}^{r's'}
                       \nonumber  \\
  &&\!\!\!+\sum_{ct}\la \overline{\Phi}_a^r\m \he \m\overline{\Phi}_c^t \ra
  \overline{x}_c^t +
  \sum_{cdtu}\la \overline{\Phi}_{a}^{r}\m \he \m\overline{\Phi}_{cd}^{tu}\ra
  \overline{x}_{cd}^{tu}\Bigg ]_{\rm linked}
  \!\!\!\!\!\!\!\!\!\!\!\!\!\m\overline{\Phi}_a^r\ra \la \Phi_0 \m
 \label{eqn2.4}
\end{eqnarray}
The wave-operator defined by equation (\ref{eqn2.4}) has only one order of
\edm as terms of the form 
$\langle \Phi_i|\edm |\overline{\Phi}_j\rangle \overline{x}$ are excluded. 
The equation of $\Omega_{_{\rm PTV}}(2)$ is 
\begin{eqnarray}
  \bigg [ \Omega_{_{\rm PTV}}(2),H_0 \bigg ]P &= &\sum_{abrs}\Bigg [ 
  \sum_{a'r'}\la\overline{\Phi}_{ab}^{rs}\m\edm\m\Phi_{a'}^{r'}\ra x_{a'}^{r'}
  + \sum_{a'b'r's'}\la \overline{\Phi}_{ab}^{rs}\m\edm\m\Phi_{a'b'}^{r's'}
  \ra x_{a'b'}^{r's'} +\sum_{ct}\la \overline{\Phi}_{ab}^{rs}\m \he \m
  \overline{\Phi}_c^t \ra \overline{x}_c^t \nonumber \\
  &&+ \sum_{cdtu}\la \overline{\Phi}_{ab}^{rs}\m \he \m\overline{\Phi}_{cd}^{tu}
  \ra \overline{x}_{cd}^{tu}\Bigg ]_{\rm linked}
  \!\!\!\!\!\!\!\!\!\!\!\!\!\!\m\overline{\Phi}_{ab}^{rs}\ra \la \Phi_0 \m
 \label{eqn2.5}
\end{eqnarray}
The term $\langle \overline{\Phi}_{ab}^{rs}|\edm|\Phi_0\rangle $ does not 
contribute as the one-body interaction Hamiltonian \edm cannot create
double excitations. The equations (\ref{eqn2.2})--(\ref{eqn2.5}) are the 
required wave-operator equations.


\subsection{Size Consistency with Connected Diagrams}

\begin{figure}[h]
  \epsfig{height=1in,width=6in,figure=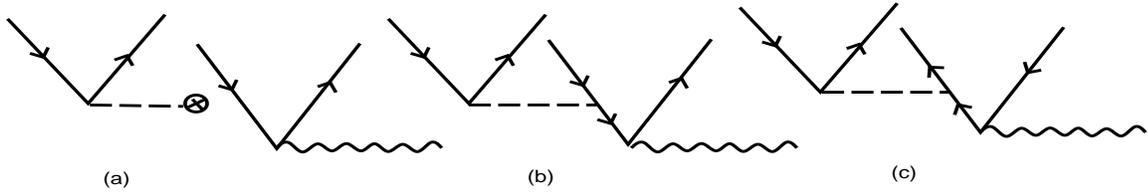} 
  \caption{The diagrammatic representations of the term $\langle 
           \Phi_{ab}^{rs}|\he |\Phi_{a'}^{r'}\rangle x_{a'}^{r'}$ }
  \label{fig3}
\end{figure}

 Consider the term $\la\Phi_{ab}^{rs}\m \he \m\Phi_{a'}^{r'}\ra x_{a'}^{r'}$ in
(\ref{eqn2.3}) the diagrams of which are given in Fig.\ref{fig3}. Among the 
diagrams (a) is linked but disconnected and remaining are connected. The 
disconnected diagrams in the wave-operator can introduce unlinked terms in the 
next iteration and hence a selection of only linked terms is required, this is 
difficult in terms of CSFs as all the contributions are combined. The other 
method of separating the wave-operator is in terms of connected cluster 
operators. It is easier to select connected terms, since it can be done 
without separating cluster operator into subcomponents. Consider the term 
$\langle\Phi_{ab}^{rs}|\he|\Phi_{a'}^{r'}\rangle x_{a'}^{r'}$ again, the 
disconnected contributions has $a'$ and $r'$ in $|\Phi_{ab}^{rs}\rangle$ and 
excitation is through the one-body part in $\he$, avoiding these terms make 
the contributions from 
$\langle\Phi_{ab}^{rs}|\he|\Phi_{a'}^{r'}\rangle x_{a'}^{r'}$ connected.
The third term in (\ref{eqn2.5}) still has disconnected terms but using the
the same method these can be removed. After these modifications all the terms 
of the cluster equations are connected. To distinguish from the linked diagram
excitation operators define the cluster-operator as $T_n$, then 
\begin{equation}
  T_n = \bigg ( \Omega(n) \bigg ) _{\rm conn} \,\,\,\, {\rm and }\,\,\,\,
  T = \sum_{n=1}^{N} T_n
\end{equation}
Taking only the linear terms of  one and two-particle cluster-operators, the
wave-operator is
\begin{equation}
  \Omega_{\rm es}=1 + T_{\rm es}(1) + T_{\rm es}(2)\,\,\,\, {\rm and }\,\,\,\,
  W = \he T .
\end{equation}
The wave-operator $\Omega_{\rm es}$ is approximated by the linear cluster
terms for the following reasons:
\begin{enumerate}
 \item The correlation introduced by $T_1^2$ is very small compared to the
       contribution from $T_2$, which represents a  large part of the 
       electron-electron correlation effect.
 \item Among the four-body cluster operators $T_2^2$ is the major contributor
       but in the present formalism this term can not be included as the 
       CSF coupling is not in particle-hole form.
 \item Though $T_1$ does not contribute significantly to the electron-electron
       correlation it is important since \edm and the dipole operators are 
       single-electron operators.
\end{enumerate}
The equation of the cluster-operator treating \he as the perturbation is 
\begin{equation}
 \bigg [ T_{\rm es}, H_0\bigg ] P = \bigg(  Q\he \Omega_{\rm es}P - 
   \chi_{\rm es}WP \bigg ) _{\rm conn}
\end{equation}
In closed-shell systems $\chi_{\rm es}W$ is always disconnected and do not
contribute to the cluster equation
\begin{equation}
  \bigg [ T_{\rm es}, H_0\bigg ] P = \bigg(  Q\he \Omega_{\rm es}P 
  \bigg ) _{\rm conn}.
\end{equation}
These are the CEPA-0 equations and do not include the EPV diagrams. The linked 
EPV diagrams should be avoided but  unlinked EPV terms should be retained. 
After suitable transformations the unlinked EPV terms can be converted into 
connected terms\cite{frantz}, thus the cluster-operator equation is
\begin{equation}
  \bigg [ T_{\rm es}, H_0\bigg ] P = \bigg(  Q\he \Omega_{\rm es} P
  \bigg )^{\rm EPV} _{\rm linked}+ \bigg ( Q\he \Omega_{\rm es}P
  \bigg)^{\rm EPO}_{\rm conn}.
\end{equation}
Where the first term is EPV and second term is non-EPV. By rearranging
\begin{equation}
  \bigg ( Q\he \Omega_{\rm es}P\bigg )^{\rm EPV}_{\rm linked} = 
  - \bigg ( Q\chi_{\rm es}WP \bigg )^{\rm EPV} .
\end{equation}
Then, the cluster-operator equations are 
\begin{equation}
  \bigg [ T_{\rm es}, H_0\bigg ] P=\bigg ( Q\he \Omega_{\rm es}P
  \bigg )^{\rm EPO}_{\rm conn} - \bigg ( Q\chi_{\rm es}WP \bigg )^{\rm EPV}
\end{equation}
The one-particle cluster operator equation is
\begin{equation}
  \bigg [ T_{\rm es}(1), H_0 \bigg ]P = \!\!\sum_{ar}\Bigg [ 
  \la \Phi_{a}^{r}\m \he \m\Phi_0\ra + \!\!\sum_{a'r'} \la \!\Phi_{a}^{r}\m \he \m
  \Phi_{a'}^{r'}\ra {\cal T}_{a'}^{r'} + \!\!\!\sum_{a'b'r's'} \!\!\!\la \Phi_{a}^{r}\m \he 
  \m \Phi_{a'b'}^{r's'} \ra {\cal T}_{a'b'}^{r's'} 
  -\!\bigg ({\cal T}^r_aW\bigg )^{\rm EPV}\Bigg ] \m\Phi_{a}^{r}\ra \la 
  \Phi_0 \m .
 \label{eqn2.6}
\end{equation}
The cluster amplitudes are denoted by ${\cal T}$ to distinguish from the
one-particle cluster amplitudes represented by $t$. Here the calculation is 
using CSFs and ${\cal T}_a^r $ is the amplitude of the cluster operator which 
excites the reference CSF $|\Phi_0\r$ to the CSF $|\Phi_a^r\r$. Similarly,
the two-particle cluster amplitudes can be defined.
\begin{equation}
  \bigg [ T_{\rm es}(2), H_0 \bigg ]P = \!\!\sum_{abrs}\Bigg [ \la \Phi_{ab}^{rs}
  \m \he \m\Phi_0\ra + \!\!\sum_{a'r'} \!\la \Phi_{ab}^{rs}\m \he \m \Phi_{a'}^{r'}
  \ra {\cal T}_{a'}^{r'} + \!\!\!\sum_{a'b'r's'} \!\!\!\la \Phi_{ab}^{rs}\m \he \m 
  \Phi_{a'b'}^{r's'} \ra {\cal T}_{a'b'}^{r's'} -\!\bigg ( {\cal T}^{rs}_{ab} W
  \bigg )^{\rm EPV} \Bigg ] \m\Phi_{ab}^{rs}\ra \la \Phi_0 \m .
 \label{eqn2.7}
\end{equation}
Similarly, the PT-violating cluster-operators $T_{_{\rm PTV}}$ are evaluated 
using the equations
\begin{eqnarray}
  \bigg [ T_{\rm es,edm}(1),H_0 \bigg ]P\!\!\! &= &\!\!\!\sum_{ar}\Bigg [ \la 
  \overline{\Phi}_a^r\m\edm\m\Phi_0\ra + \sum_{a'r'}\la \overline{\Phi}_a^r\m
  \edm\m\Phi_{a'}^{r'}\ra {\cal T}_{a'}^{r'} + \!\!\sum_{a'b'r's'}\!\!\la
  \overline{\Phi}_a^r\m\edm\m\Phi_{a'b'}^{r's'}\ra {\cal T}_{a'b'}^{r's'}
                       \nonumber  \\
  &&\!\!\!\!+\sum_{ct}\!\la \overline{\Phi}_a^r\m \he \m\overline{\Phi}_c^t \ra
  \overline{\cal T}_c^t + \!\!\sum_{cdtu}\la \overline{\Phi}_{a}^{r}\m \he \m
  \overline{\Phi}_{cd}^{tu}\ra\overline{\cal T}_{cd}^{tu} -\!\! \bigg ( 
  \overline{t}^r_a W \!\bigg )^{\rm EPV} \Bigg ]\m\overline{\Phi}_a^r\ra\la
  \Phi_0\m
 \label{eqn2.8}
\end{eqnarray}
and 
\begin{eqnarray}
  \!\!\!\!\!\!\!\!\!\!\!\!\!\!\!\!\!\!\!\!
  \!\!\!\!\!\!\!\!\!\!\!\!\!\!\!\!\!\!\!\!
  \bigg [ T_{\rm es,edm}(2),H_0 \bigg ]P \!\!&= &\!\!\sum_{abrs}\Bigg [ 
  \sum_{a'b'r's'}\la \overline{\Phi}_{ab}^{rs}\m\edm\m\Phi_{a'b'}^{r's'}
  \ra {\cal T}_{a'b'}^{r's'}
  +\sum_{ct}\la \overline{\Phi}_{ab}^{rs}\m \he \m\overline{\Phi}_c^t \ra
  \overline{t}_c^t  \nonumber \\
  &&+\,\sum_{cdtu}\la \overline{\Phi}_{ab}^{rs}\m \he \m\overline{\Phi}_{cd}^{tu}
  \ra \overline{\cal T}_{cd}^{tu} -\bigg ( \overline{\cal T}^{rs}_{ab} W
  \bigg ) ^{\rm EPV} \Bigg ] \m\overline{\Phi}_{ab}^{rs}\ra \la \Phi_0 \m
 \label{eqn2.9}
\end{eqnarray}
Using the cluster-operators 
\begin{equation}
  \Omega_{\rm es,edm} = \Omega_{\rm es,edm}(1)  +  \Omega_{\rm es,edm}(2) .
\end{equation}
The atomic EDM $d_a$ can be calculated using the operators as
\begin{equation}
  d_a = \la \Phi_0\m\vec{D}_{\rm eff} \m\Phi_o\ra ,
\end{equation}
where $\vec{D}_{\rm eff}=\Omega^{\dagger}_{\rm es}\vec{D}\Omega_{\rm es,edm}
+ \Omega^{\dagger}_{\rm es,edm}\vec{D} \Omega_{\rm es}$ is same as 
before except that the wave-operator is now in terms of connected clusters. The
cluster equations are similar to the CEPA-2 
equations\cite{kutzelnigg,ahlrichs}, the explicit CEPA-0 equations are same as 
these equations without the EPV terms.


\section{Analysis of the Cluster Equations}

 The  cluster based formalism is not an order by order formalism but an 
iterative scheme where the Bloch equation is defined in orders of
excitations rather than the orders of perturbation. It is possible to separate 
the contributions from various many-body effects cleanly using Moller-Plesset 
partitioning and an added advantage is the one-to-one correspondence with the 
diagrams using Goldstone evaluation rules. Such an approach has been 
used\cite{blundell} to calculate parity non-conservation in atomic cesium to 
very high accuracy. The E-N partitioning mixes the contributions, but it has 
the  advantage of capturing the static correlation effects very effectively. 
The CEPA-0 equations are exactly the linearized singles and doubles coupled 
cluster equations, the CEPA-2 equations includes a class of non-linear terms.


\subsection{The Singly Excited Amplitude Cluster Equation}

\begin{figure}[h]
  \epsfig{height=3in,width=\textwidth,figure=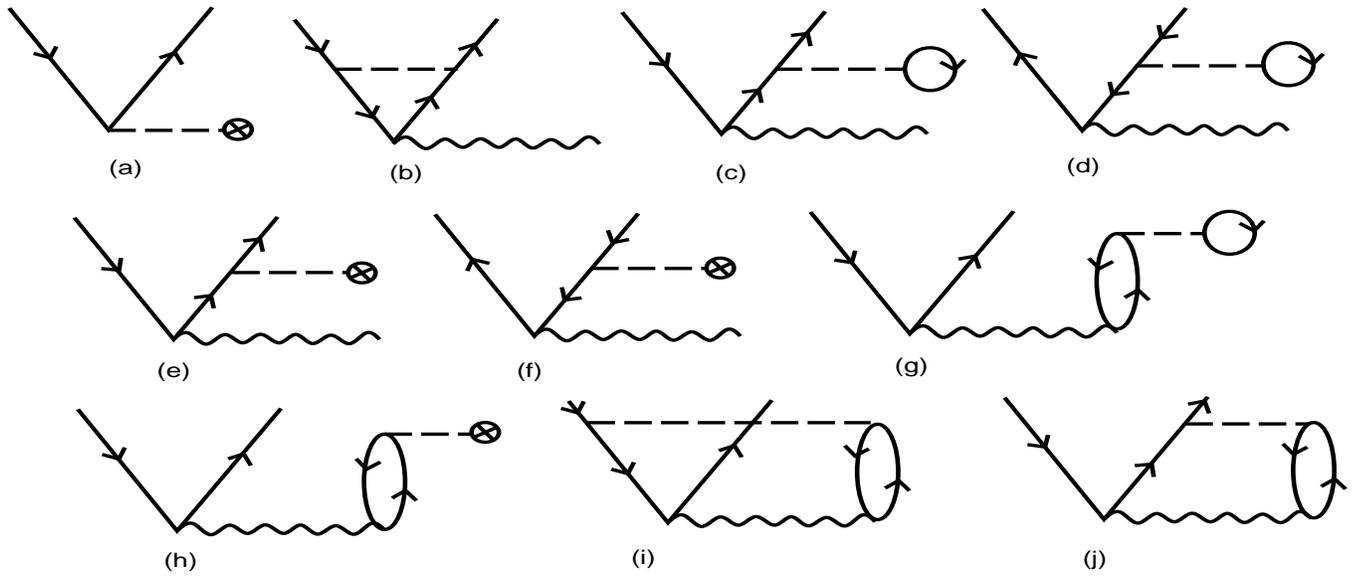}
  \caption{Diagrams that contributes to the single-excitation cluster 
           amplitude.}
  \label{fig4}
\end{figure}

The diagrammatic representation of the principal terms in (\ref{eqn2.6}) are 
shown in Fig.\ref{fig4}. The contribution from each of the diagrams are as 
described: 
\begin{enumerate}
  \item Diagram (a) contributes to the  first term 
        $\l \Phi_a^r|H_{\rm es}|\Phi_0\r$ of the cluster equation and
        is independent of any cluster amplitudes. It is an important term as 
        the iteration proceeds from this term.
  \item The diagrams (b),(c),(d),(e) and (f) contribute to the second term
        $\l \Phi_a^r|H_{\rm es}|\Phi_{a'}^{r'}\r {\cal T}^{r'}_{a'}$ and
        the $\he$ matrix element is coupled with the single excitation
        cluster amplitude. These start contributing from the second iteration,
        where the cluster amplitude ${\cal T}_{a'}^{r'}$ is just the matrix 
        element $\l \Phi_{a'}^{r'}|\he|\Phi_0\r$ in the first iteration.
  \item Diagrams (g),(h),(i) and (j) contribute to the third term $\l 
        \Phi_a^r|\he|\Phi_{a'b'}^{r's'}\r {\cal T}^{r's'}_{a'b'}$ and  
        couple the double excitation cluster amplitude with the $\he$
        matrix element. Similar to the second term, these diagrams start
        contributing from the second iteration.
\end{enumerate}

Consider the diagrams (c) and (e), though they resemble the Hartree-Fock 
potential scattering diagram these are very different. Consider the the 
bubble part of the diagram (c), it is summed over the occupied orbitals common 
to both the initial and the final CSFs in the matrix element of $\he$. An 
example in Yb is if the initial and final CSFs are 
$|\Phi_{a'b'}^{r's'}\r =|7s^2\r $ and $|\Phi_{ab}^{rs}\r = |7s8s\r$ 
respectively, the bubble part in (c) has all the occupied orbitals except the 
$6s$ orbital. This is because both the CSFs do not have $6s$, where as in the 
Hartree-Fock scattering diagram the bubble should have contribution from all 
the occupied orbitals. A similar description is true of diagrams (d) and (f) 
too.

 The term $({\cal T}_a^rW)^{\rm EPV}$ picks up the effect of the non-linear
terms  $T_{es}(1)^2$ and $T_{es}(1)T_{es}(2)$, which have $T_{es}(1)$ 
amplitudes. This implies that the wave-operator assumes the form
\begin{equation}
  \Omega_{es} = 1+ T_{\rm es}(1) + T_{\rm es}(2) +\bigg [
  T_{\rm es}(1)T_{\rm es}(1) + T_{\rm es}(1)T_{\rm es}(2) +
  T_{\rm es}(2)T_{\rm es}(2) \bigg ] ^{\rm EPV}
 \label{eqn3.1}
\end{equation}
Terms which are not included in the single excitation cluster amplitude
equation are 
\begin{equation}
  \bigg [ T_{\rm es}(1) T_{\rm es}(1) + T_{\rm es}(1)T_{\rm es}(2) + 
  T_{\rm es}(2)T_{\rm es}(2) \bigg ]^ {\rm EPO}.
\end{equation}
 A later section describes the method to choose EPV terms from the 
renormalization part.


\subsection{The Doubly Excited Cluster Amplitude Equation}

\begin{figure}[h]
  \epsfig{height=2.5in,width=\textwidth,figure=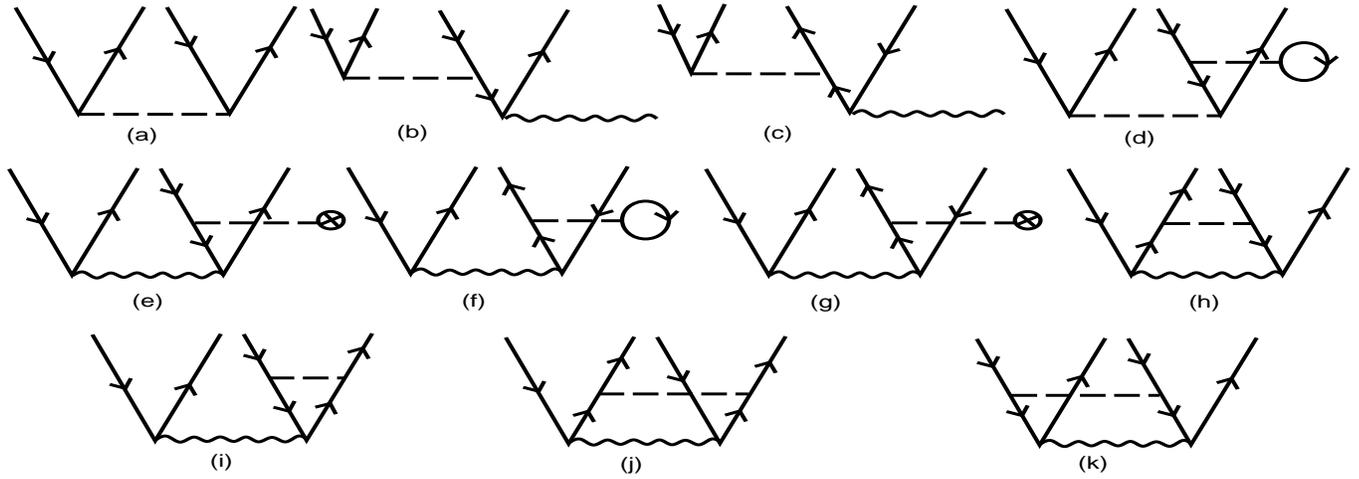}
  \caption{The diagrams for the terms in doubly-excited cluster operator 
           equation.}
  \label{fig5}
\end{figure}

 The diagrams of the principal terms of the double excitation cluster amplitude
(\ref{eqn2.7}) are shown in Fig.\ref{fig5}. The first term is similar to that 
of the single excitation cluster amplitude equation. The second term can be 
separated as
\begin{eqnarray}
  \sum_{a'r'}\la \Phi_{ab}^{rs}\m \he\m\Phi_{a'}^{r'}\ra {\cal T}_{a'}^{r'} &=& 
  \la \Phi_{ab}^{rs}\m \he\m\Phi_{a}^{r}\ra {\cal T}_{a}^{r} + \sum_{a'\ne a,b}
  \la \Phi_{ab}^{rs}\m \he\m\Phi_{a'}^{r}\ra {\cal T}_{a'}^{r} + 
  \sum_{r'\ne r,s}\la \Phi_{ab}^{rs}\m \he\m\Phi_{a}^{r'}\ra {\cal T}_{a}^{r'}
                     \nonumber \\
 &&+ \sum_{a'\ne a,b}\la \Phi_{ab}^{rs}\m \he\m\Phi_{a'}^{s}\ra {\cal T}_{a'}^{s}
  + \sum_{r'\ne r,s}\la \Phi_{ab}^{rs}\m \he\m\Phi_{b}^{r'}\ra {\cal T}_{b}^{r'}
       \nonumber 
\end{eqnarray}
The first  term on the right hand side has connected as well as disconnected 
terms, from which only the connected terms should be retained. The remaining 
terms are connected and hence linked too since the conditions  
$a'\ne a,b; r' \ne r,s $ exclude the disconnected terms. Then 
\begin{equation}
  \sum_{a'r'}\la \Phi_{ab}^{rs}\m \he\m\Phi_{a'}^{r'}\ra {\cal T}_{a'}^{r'} = 
  \!\!\la \Phi_{ab}^{rs}\m \he\m\Phi_{a}^{r}\ra {\cal T}_{a}^{r} 
  \!\!\!+\!\!\! \sum_{r'\ne r,s}\!\!\!\bigg (\delta_{a'a} + \delta_{a'b} \bigg) 
  \la \Phi_{ab}^{rs}\m \he\m\Phi_{a'}^{r'}\ra {\cal T}_{a'}^{r'}
  \!\!\! + \!\!\!\sum_{a'\ne a,b}\!\!\!\bigg (\delta_{r'r} + \delta_{r's} 
  \bigg )\la \Phi_{ab}^{rs}\m \he\m\Phi_{a'}^{r'} \ra {\cal T}_{a'}^{r'} 
\end{equation}
Similarly, the third term can be expanded to 
\begin{eqnarray}
  \sum_{a'b'r's'}\!\!\!\la \Phi_{ab}^{rs}\m \he\m\Phi_{a'b'}^{r's'}\ra 
  {\cal T}_{a'b'}^{r's'}\!\!\! &= & \!\!\!\!\!
  \sum_{b'\ne a,b}\!\!\!\sum_{s'\ne r,s}\la \Phi_{ab}^{rs}\m \he\m
  \Phi_{ab'}^{rs'}\ra {\cal T}_{ab'}^{rs'}  + \!\!\! 
  \sum_{r'\ne r,s}\Bigg [ \la \Phi_{ab}^{rs}\m \he\m\Phi_{ab}^{r's}\ra 
  {\cal T}_{ab}^{r's} +\!\!\!\sum_{s'\ne r,s}\!\!\!\la \Phi_{ab}^{rs}\m \he\m
  \Phi_{ab}^{r's'} \ra {\cal T}_{ab}^{r's'}\Bigg ] \nonumber \\
  &&\!\!\!+\!\!\!
  \sum_{a'\ne a,b}\Bigg [ \la \Phi_{ab}^{rs}\m \he\m \Phi_{a'b}^{rs} \ra 
  {\cal T}_{a'b}^{rs} +\!\!\!\sum_{b'\ne a,b}\!\!\!\la \Phi_{ab}^{rs}\m \he\m
  \Phi_{a'b'}^{rs}\ra {\cal T}_{a'b'}^{rs}\Bigg ], \nonumber 
\end{eqnarray}
where all the terms are connected. The triply and quadruply excited terms 
are excluded. Each of the diagrams has an exchange diagram too. Terms 
corresponding to each of the diagrams are:
\begin{enumerate}
  \item  Diagram (a) correspond to the first term in the cluster equation 
         and has no dependence on any of the cluster amplitude.
  \item $\l\Phi_{ab}^{rs}|\he|\Phi_{a}^{r}\r {\cal T}_{a}^{r}$ 
        contribute to diagrams (b) and (c). The final CSF in this term 
        has a hole-particle pair in common with the cluster amplitude. 
  \item $(\delta_{a'a}+\delta_{a'b})\l\Phi_{ab}^{rs}|\he|\Phi_{a'}^{r'}
        \r {\cal T}_{a'}^{r'}$ contribute to diagram (c). Though the topology
        of the diagram is same as that of 
        $\l\Phi_{ab}^{rs}|\he|\Phi_{a}^{r}\r {\cal T}_{a}^{r}$, it is an EPO 
        diagram. The $\he$ interaction changes the state of the particle 
        and picks up a part of core-virtual correlation effect, which can be 
        identified as core-polarization.
  \item $(\delta_{r'r}+\delta_{r's})\l\Phi_{ab}^{rs}|\he|\Phi_{a'}^{r'}
        \r {\cal T}_{a'}^{r'}$ contribute to diagram (b). This also has similar 
        topology with $\l\Phi_{ab}^{rs}|\he|\Phi_{a}^{r}\r {\cal T}_{a}^{r}$ 
        but is again an EPO diagram, where there is a change of the hole state 
        and correspond to core-core correlation effect. 
  \item $\l\Phi_{ab}^{rs}|\he|\Phi_{ab'}^{rs'}\r {\cal T}_{ab'}^{rs'}$ 
        contribute to diagram (h) and (i). These are EPO diagrams where a 
        hole-particle change to another hole-particle pair. These 
        contribute to the core-virtual correlation effects.
  \item The term $\l\Phi_{ab}^{rs}|\he|\Phi_{ab}^{r's}\r {\cal T}_{ab}^{r's}$ 
        contribute to diagrams (f) and (g). These are EPO diagrams where one
        of the particle states in ${\cal T}_{ab}^{r's}$ is excited to 
        another particle state. This can also contribute to EPV diagrams 
        of the first kind, if it is a hole-line EPV diagram then it will 
        correspond to (h) and (i) and if it is particle line EPV then diagram 
        (j).
  \item $\l\Phi_{ab}^{rs}|\he|\Phi_{ab}^{r's'}\r {\cal T}_{ab}^{r's'}$ 
        contribute to diagram (j). This is a double excitation where the 
        particle states from the cluster amplitude ${\cal T}_{ab}^{r's'}$ are 
        excited to different particle states but the hole states remain
        intact. These terms capture the virtual-virtual correlation effects.
  \item $\l\Phi_{ab}^{rs}|\he|\Phi_{a'b}^{rs}\r {\cal T}_{a'b}^{rs}$ 
        contribute to diagram (d) and (e). These diagrams correspond to a
        change of the hole state and are EPO diagrams which capture the 
        single-body hole-hole interaction component. This term can also 
        contribute EPV diagrams, the hole line EPV diagram arising from this 
        term is (k) and the particle line EPV diagrams are (h) and (i).
  \item $\l\Phi_{ab}^{rs}|\he|\Phi_{a'b'}^{rs}\r {\cal T}_{a'b'}^{rs}$ 
        contribute to (j) and is a hole-hole correlation term.  
\end{enumerate}
Thus terms in the cluster equation contribute to different physical effects. 
So far only the first three terms in the cluster equation have been considered.
The last term in the doubly excited cluster amplitude  
$({\cal T}^{rs}_{ab} W )^{\rm EPV}$ is a renormalization term. It picks up a 
set of terms non-linear in cluster amplitude 
\begin{equation}
  T_{\rm es}(1)T_{\rm es}(2) + T_{\rm es}(2) T_{\rm es}(2).
\end{equation}
These pick up a class of EPV terms which are non-linear in cluster amplitudes.
Consider the expression of $W$
\begin{equation}
  W = P\he \bigg (T_{\rm es}(1) + T_{\rm es}(2)\bigg ) P .
\end{equation}
The term $T_{\rm es}(1)T_{\rm es}(2)$ is picked up through 
$P\he T_{\rm es}(1)P$ in $W$, which implies that 
$({\cal T}_{ab}^{rs}P\he T_{\rm es}(1))^{\rm EPV}$ can have one hole(particle)
EPV line or a pair of hole-particle EPV lines. Whereas in the single 
excitation cluster amplitude equation, the contribution from
$T_{\rm es}(1)T_{\rm es}(2)$ is captured through the term
$({\cal T}_{a}^{r}P\he T_{\rm es}(2))^{\rm EPV}$ in 
$({\cal T}_{a}^{r}W)^{\rm EPV}$. But the number of EPV hole-lines or EPV 
particle-lines are the same in both. In general the number of EPV hole-lines 
and particle-lines in $(T_{\rm es}(n)P\he T_{\rm es}(m)P)^{\rm EPV}$ is limited
by the $\he$ if $m,n > 2$ and by the cluster amplitudes if $m< 2$ or $n < 2$. 
Although $T_{\rm es}(1)T_{\rm es}(2)$ is included in the single  as well as 
the double excitation cluster amplitudes the topology of the diagrammatic 
representations are different. Diagrams from $({\cal T}_a^rW)^{\rm EPV}$ has 
only a pair of hole-particle lines where as $({\cal T}_{ab}^{rs}W)^{\rm EPV}$ 
has two pairs of hole-particle lines.


\subsection{Selection of EPV Terms and Connected Terms}

  The terms linear in cluster amplitude cannot 
violate Pauli exclusion principal. It is possible only when there are CSFs 
which are EPV, which is not possible. The EPV diagrams arise from 
renormalization terms, which are non-linear in cluster amplitude . The diagrams 
representing the renormalization terms are the cluster diagrams multiplied
by the energy diagrams and hence unlinked but these can suitably be rearranged 
to yield connected diagrams.
 
 To select the EPV terms all the orbitals are tagged with labels which are 
prime numbers. The CSFs are also assigned a number which is the product of the 
prime numbers corresponding to the labels of the holes and particles of the 
CSFs. Consider a doubly excited CSF $|\Phi_{ab}^{rs}\rangle$, the prime 
numbers $n_a, n_b, n_r\mbox{\rm and } n_s$ be the labels of the hole and 
particle states and ${\cal N}_{ab}^{rs}$ their product, these five numbers 
identify the CSF. However, only three if the CSF is singly excited. To maintain
consistency the remaining two indices are filled with another prime number not 
used in labeling the orbitals, let this number be $N_P$ but set the 
corresponding multiplying factor in $n$ as unity. According to this scheme, the
ground/reference state of Yb is identified by  
$N_P, N_p, N_p, N_p \mbox{\rm and} 1$. Similarly, labels are also given to the 
cluster amplitudes. 

 The terms to retain from $({\cal T}_a^rW)^{\rm EPV}$ are those having CSFs 
in $W$ identified by ${\cal N}$ which can be divided by one or more of 
the numbers identifying  hole(particle) of the cluster amplitude ${\cal T}$. 
The number of possible division is the number of common hole/particle lines 
between ${\cal T}$ and $W$. During the selection process division by $N_p$ 
should be discarded as this does not represent any hole of particle states and 
this is automatically achieved as the corresponding multiplying factor in $n$ 
is unity. The advantage of this scheme is that it reduces the number of 
operations required in the selection process.

 The term $\l\Phi_{ab}^{rs}|\he|\Phi_{a'}^{r'}\r {\cal T}_{a'}^{r'}$ of the 
doubly excited cluster amplitude equation has disconnected components if both 
the hole and particle states in the initial CSF are present in the final CSF. 
These are discarded and only the connected components are chosen. 
This can be implemented while calculating the matrix elements of $\he$.
During the matrix element calculation the total number of hole states of the
initial and final CSFs is calculated. If this is equal to three then these 
contribute to $\l\Phi_{ab}^{rs}|\he|\Phi_{a'}^{r'}\r {\cal T}_{a'}^{r'}$. In 
the next step the connected component is chosen by selecting the 
$\he$ matrix element which has $r'$. Then the doubly excited cluster amplitude 
equation has only connected components. However the discarded components are 
disconnected but linked. 


\section{The Configuration Space Considered}
 The configuration space is spanned by the CSFs constructed from the 
$V^{\rm N-1}$ orbitals. The CSFs are generated by single or double 
excitations from the occupied orbitals to the bound and the continuum virtual 
orbitals in all possible ways such that it yields the required final angular 
momentum. For the single reference MBPT, the reference CSF of Yb is $|6s^2\r$, 
which be referred as $|\Phi_0\r$. Hence the occupied orbitals
are $(1$\,-$6)s$,$(2$\,-$5)p*$, $(2$\,-$5)p$,$(3$\,-$4)d*$, $(3$\,-$4)d$,$4f*$ 
and $4f$ respectively. The CSF space generated is not a complete 
active space but complete for the single and double excitations from the most 
important outer occupied orbitals within the orbital orbital space.  
The occupied-orbital shells of the configurations that has been considered are
\begin{tabbing}
single excitation:\hspace*{.4cm}\=$\m 4f\!*^64f^86s\ra$, $\m 4f\!*^56s^{2}\ra$ 
                    $\m 4f^76s^{2}\ra $, $\m 5p\!*^{1}6s^{2}\ra$  
                    $\m 5p^{3}6s^{2}\ra$ and $\m 5s^16s^2\ra$ \\
double excitation:\>$\m 4f\!*^{6}4f^{8}\ra$, $\m 4f\!\!*^{5}6s\ra$
                    $\m 4f^{7}6s\ra$, $\m 4f\!\!*^{4}6s^{2}\ra$,
                    $\m 4f^{6}6s^{2}\ra$, $\m 4f\!\!*^54f^76s^2\ra$
                    $\m 5p\!*^16s\ra$,$\m 5p^36s\ra$,\\
                  \>$\m 5p\!*^14f^56s^2\ra$, $\m 5p^34f\!\!*^56s^2\ra$, 
                    $\m 5p\!*^34f^56s^2\ra$, $\m 5p^34f^56s^2\ra$,
                    $\m 5s^16s\ra$,$\m 5s^14f\!\!*^56s^2\ra$, 
                    $\m 5s^14f^76s^2\ra$, \\
                  \>$\m 5s^15p\!*^16s^2\ra$  and $\m 5s^15p^36s^2\ra$. \\
\end{tabbing}
The remaining electrons are distrubuted among the virtual orbitals in all 
possible ways. From all the CSFs only the $J\!=\!0$ even parity and $J\!=\!1$
odd parity CSFs are chosen. The number of non-relativistic CSFs generated are 
given in Table\ref{table1}. Although not included in the table, CSFs with 
excitations from $5s$ are also included in the CSF space.

 The total number of odd and even parity CSFs with bound virtual orbitals are 
9930 and 17087 respectively. The modulus of the EN-partitioned energies of the 
CSFs--the diagonal Hamiltonian matrix elements--are as shown in the histograms
Fig.4. The two histograms are plotted such the the lowest $|E|$ is shifted to 
zero and the the range between the lowest and the highest are divided into ten 
units. The zero on $|E|$ axis are 14064.9531 and  14065.0068 hartrees for the 
even and odd parity CSFs respectively. Similarly, the highest $|E|$ are 
14067.6720 and 14067.5996 hartrees respectively. From the histogram, the number 
of CSFs with low and high energies are less whereas the number of the 
configurations that can give the intermediate energy are large. As a result
the perturbation series converges fast as only a few of the configurations are 
quite close to $|\Phi_0\r$ and the energy separation with the rest of the CSFs 
is quite large.

   Number of odd parity CSFs in the intermediate energy is more than the even 
parity as the odd parity configuration space can have many possible 
intermediate couplings to give $J\!=\!1$, to limit the number of CSFs within the
memory limitations a selection of CSFs is done. The double excitations to
$d$ and $f$ symmetries above the energy of the converged orbitals are not 
included. Another constraint on the choice of configurations is: there 
shouldn't be more than four open shells in the non-relativistic notation and 
eight in the relativistic form, choosing only singly and doubly excited 
configurations satisfies this condition for a closed-shell atom like Yb. This 
constraint is due to the angular co-efficient computation program.


\section{Results}

 The atomic Yb EDM is calculated using the Bloch equation based MBPT and 
the size-consistent CEPA equations. As a part of the calculation the ground
state energy is also computed. The results with different methods are given
in the following sections.


\subsection{Bloch Equation Based MBPT}
\subsubsection{Calculation of $\Omega_{\rm es}$ and $E_0$ }

  Using the wave-operator $\Omega_{\rm es}$ the ground state wave-function 
$|\Psi_0\r $ and energy $E_0$ are
\begin{equation}
  \m \Psi_0 \ra = \frac{\Omega_{\rm es}\m \Phi_0\ra}
  {\la \Phi_0\m\Omega_{\rm es}^{\dagger}\Omega_{\rm es}\m \Phi_0\ra }
  \;\;\;\;\;{\rm and  } \;\;\;\;\;
  E_0 = \la \Phi_0\m \he\Omega_{\rm es}\m \Phi_0\ra .
\end{equation}
The denominator in the expression of $|\Psi_0\rangle$ is the normalization
factor. The first order energy correction of the ground state is zero 
Since the calculations are using the EN-partitioned  Hamiltonian. 

The  even parity CSF space of the calculation is spanned by 9930 
CSFs, of which core part of first 4435,  4436-9094 and remaining CSFs are 
$|4f^{14}6s\r , |4f^{14}\r , |4f^{13}6s^2\r |4f^{13}6s\r$ and 
$|4f^{12}6s^2\r$, and $|5p^54f^{14}6s^2\r$, $|5p^54f^{14}6s\r$ and  
$|5p^54f^{13}\r$, and  $|5s5p^64f^{14}6s^2\r$ respectively. Using this set of 
CSFs the ground state wave-function is 
\begin{eqnarray}
  \m \Psi_0 \ra &=& 0.9251\;23\m 6s^2\ra + 0.1172\;17\m 6p*\!^2 \ra 
  + 0.1169\;21 \m6s7s \ra + 0.0996\;76\m 6p\!^2 \ra 
                          \nonumber \\
  &&-0.0600\;60\m 5d^2\ra + 0.0568\;61\m 6p7p\ra -0.0497\;43\m 5d\!*^2 \ra 
  +0.0480\;54\m 6p\!*7p\!*\ra  
                          \nonumber \\
  &&+0.0443\;18\m 6p8p \ra -0.0442\;15\m 6s8s\ra-0.0320\;73\m 5d6d\ra +\ldots .
                          \nonumber 
\end{eqnarray}
Where only the ten important CSFs are given explicitly  and the normalization 
constant is $1.0809\;37$. As expected, the most important CSFs in $|\Psi_0\r$ 
are doubly excited except for $|6s7s\r$, which is a singly excited CSF and does
not interact very strongly with the ground state CSF $|6s^2\r$ but contributes 
significantly by correlation through other CSFs when the residual interaction 
is taken to higher orders. Values of $E_0$  with increasing size of the  even 
CSF space is given in Table\ref{table2}. CSFs are added to the calculation in 
sequence of excitations from the deeper core orbitals. As the size of the CSF 
space is increased more complicated many-body effects are included in the 
computation.

  An important quantity that can be extracted from Table\ref{table2} is 
the change in $E_0$.  Define the correlation energy $\Delta E_0$ as the energy 
difference between the CSF energy of $|6s^2\r$  and the energy calculated 
using the CSFs in the even-parity CSF space. From the plot of $\Delta E_0$
in Fig\ref{fig7}a it is evident that the change in the ground state ASF energy 
is not uniform but in steps interrupted by regions of very minimal changes. 
Most significant changes of $\Delta E_0$ occur while increasing the CSFs from 
100 to 500, from 1000 to 2000  and from 6435 to 7435 CSFs respectively. These 
changes are not the combined effect of all the CSFs added but due to a few 
important ones. The largest change of $\Delta E_0$ is while increasing from
1000 to 2000 CSFs which corresponds to contribution from the core 
configuration $|4f^{14}\r$, that is double excitation from the $6s$ orbital 
shell. The remaining two are due to the core configurations $|4f^{13}6s\r$ 
and $|5p^56s\r$. Each contribute $-0.0041\;82$, $-0.0179\;26$ and 
$-0.0045\;07$ hartrees respectively, the combined effect adds to 81.20\% of 
the total correlation energy  $-0.0304\;50$ hartrees. From these it can be 
concluded that, the most important CSFs contributing to the correlation energy
have core configurations $|4f^{13}6s\r$, $|4f^{14}\r$ and $|5p^56s\r$. As to 
be expected the doubly excited CSFs are most important to capture the 
correlation effects and the low lying double excitations from $6s$ orbital
shell has the most significant contribution to correlation energy, it 
contributes 54.74\% of the correlation energy.

The plot (b) in Fig:\ref{fig7} indicates the need to include $\he$ to high 
orders to capture the correlation effects accurately. From the graph the 
correlation effect due to two order of $\he$ is $-0.0436\;033$ hartrees and
decreases in magnitude monotonically till fourth order to $-0.0306\;67$hartrees
but increases in the fourth order to $-0.0343\;38$hartrees. This trend of
oscillation about the final value of $\Delta E_0$ continues till convergence. 
The cycle of the oscillation has a period of four orders, that is in four 
orders it goes to the same side of the final value of $\Delta E_0$ and the 
amplitude of the oscillation decreases with each cycle. Over all, the value 
of $\Delta E_0$ behaves like a damped oscillator with a cycle of four orders. 
If $E_0$ is calculated by truncating the perturbation to the  first few orders
where the amplitude of oscillation is quite significant the value of 
$\Delta E_0$ can be erroneous.

 The Fig.\ref{fig8} shows the trend of wave-operator convergence. The first 
graph Fig.\ref{fig8}a is the value of the convergence criteria plotted against 
the order of perturbation and second graph Fig.\ref{fig8}b is $\log_{10}$ of 
the convergence criteria plotted against the order of perturbation. From the 
first graph it is evident that the convergence criteria begins with a small
value but as shown in the second graph in terms of order of magnitude, the 
convergence is not so fast.  The convergence is monotonic with very regular 
fluctuations. The wave-operator $\Omega_{\rm es}$ is stored in an order by 
order sequence. These are accessed as and when required  during the calculation
of $\Omega_{\rm es,edm}$.


\subsubsection{Calculation of $\Omega_{\rm es,edm}$ and $d_a$ }

  The wave-operator $\Omega_{\rm es,edm}$ is calculated introducing the 
interaction Hamiltonian \edm and adding opposite parity CSFs to the CSF space. 
Once \edm \ is applied to the wave-operator $\Omega_{\rm es}$, it maps onto the 
odd-parity component of the CSF space and can never be mapped back to the 
even-parity space as \edm \ is treated to first order only. This is followed by 
a sequence of residual coulomb interaction $\he$, which accounts for the 
correlation effects within the odd-parity sub-space. In sum total it is a 
sequence of perturbations applied to the ground state CSF $|6s^2\r$, where \edm 
is sandwiched between all possible arrangements of $\he$. After the wave-operator
$\Omega_{\rm es.edm}$ is calculated, the mixed parity ground state wave-function 
$|\widetilde{\Psi}_0\r$ can be written in terms of the wave-operators 
$\Omega_{\rm es}$ and $\Omega_{\rm es,edm}$ as
\begin{equation}
  \m \widetilde{\Psi}_0 \ra = \m \Psi_0\ra + \m \Psi_{\rm corr}^0\ra =
  \bigg (\Omega_{\rm es} + \Omega_{\rm es,edm}\bigg ) \m \Phi_0 \ra
\end{equation}
The value of $d_a$ can be calculated using the expression
\begin{equation}
  d_a = \la \widetilde{\Psi}_0 \m \vec{D} \m \widetilde{\Psi}_0\ra
  =2\la \Phi_0 \m \Omega_{\rm es}^{\dagger}\vec{D} \Omega_{\rm es,edm}\m \Phi_0 \ra
\end{equation}
Choosing the odd parity CSF as mentioned in Sec. V the correction to the ground 
state $|\Psi_0\r$ from the opposite parity sub-space due to \edm is 
\begin{eqnarray}
  \m \Psi_{\rm corr}^0 \ra\!\! &= &\!{\cal A}\bigg (-55.1403\;73 \m 6s 6p\!* 
  \!\ra -17.3681\;84\m 6s 7p\!*\!\ra + 10.8231\;86 \m 6s6p \ra 
                                \nonumber  \\
  &&\!\!\!-9.5064\;23 \m 5p\!*7s\ra -9.2841\;00 \m 5s 6p\!*\! \ra 
  +7.8322\;72\m 6s 8p\!*\!\ra +7.1950\;56 \m 6p\!* 5d*\!\ra 
                                \nonumber \\
  &&\!\!\!+5.5464\;22\m 6s 9p\!*\!\ra+5.5189\;58 \m 6p5d\!*\! \ra 
  +5.1246\;77\m 5p\!*8s\!\ra +\ldots \bigg ) .
   \nonumber
\end{eqnarray}
Where ${\cal A}= \sqrt{2}C_T\sigma_NG_{_F}$ and only the first ten important 
CSFs are listed. The coefficients are much larger than unity but these should be 
scaled by the parameter $ \sqrt{2}G_{_F}$. The product of the 
coupling constant $C_T$  and nuclear spin $\sigma_N$ is retained as a 
parameter and $C_T$ can be estimated after combining with the experimental 
results. The above expression for $|\Psi_{\rm corr}^0\r$ shows that:
\begin{enumerate}
  \item Like in the lowest order single-particle calculation, the coefficients 
        of the CSFs $|6snp\!*\r$ flip sign for $n\!\ge\!8$. 
  \item Most of the important CSFs are singly excited with respect to the ground
        state CSF $|6s^2\r$, which is to be expected since \edm \ is a single
        particle interaction Hamiltonian.
  \item Singly excited configurations like $|6s6p\r$ can contribute through
        three possible many-body routes. First \edm \ excites $|6s^2\r$ to 
        $|6snp\!*\r$ then a sequence of $\he$ connects it to $|6s6p\r$, second 
        a sequence of $\he$ connects $|6s^2\r$ to $|6p\!*6p\r$ and \edm takes
        it to $|6s6p\r$ and third a sequence of $\he$ takes $|6s^2\r$ to a 
        CSF $|\Phi_i\r$ which connects to $|\Phi_j\r$ via \edm and another 
        sequence of $\he$ connects it to $|6s6p\r$. Although $|6s6p\r$
        cannot connect directly to the ground state $|6s^2\r$ as $6p$ is 
        close to zero within the nucleus, it is the third most important CSF 
        which contributes to $|\Psi_{\rm corr}^0 \r $. This demonstrates the 
        importance of many-body effects.
  \item The two most important doubly excited odd-parity CSFs for the evaluation
        of EDM are $|6p\!*5d\!\r$ and $|6p5d\!*\r$. More interesting is the 
        second as both the virtual orbitals involved cannot contribute to 
        the \edm \ matrix elements. Among the possible many-body routes which 
        can contribute to the co-efficient of $|6p5d\!*\r$ one possibility is 
        through the deeper occupied orbitals $5p*$ and $5s$, which would 
        contribute to core polarization effects.
\end{enumerate}
The value of $d_a$ is calculated using different sets of CSFs, where the number 
of either the even or odd CSFs are fixed to the maximum allowed and then 
increase the number of CSFs in the opposite parity CSF space. The results of 
such a sequence of calculations is given in the Table \ref{table3}. If the 
previous sequence of calculation shows the importance of occupied orbitals 
in the whole CSF space, these two sequences demonstrates the significance of the 
occupied orbitals in CSF sub-space of each parity.

  Like the correlation energy $\Delta E_0$ there is a significant change of
$d_a$ when CSFs with double excitations from $6s$ are included. To appreciate 
the change better the values of $d_a$ in the two sequence are  plotted in 
Fig.\ref{fig8}. Consider the sequence where the number of even-parity CSFs is 
fixed, $d_a$ increase with the number of the odd-parity CSFs, which implies  
that there are no appreciable cancellations due to the odd parity CSFs added. In
the second sequence where the number of odd-parity CSFs is fixed, $d_a$ 
decreases as the number of even-parity CSFs is increased. 

  In both the sequence there is a significant change in $d_a$ when CSFs with 
double excitation from $6s$ orbital are added. But, the changes are different in
sign, in the even CSF space the inclusion of CSFs with double excitation from 
$6s$ increases the value of $d_a$ where as in the odd-parity CSFs it decreases. 
Consider the expression for EDM it can be expanded as
\begin{equation}
  d_a = 2\bigg (\la \Phi_0 \m\vec{D} \Omega_{\rm es,edm}\m \Phi_0 \ra
  +\sum_n\la \Phi_0 \m \Omega_{\rm es}^{\dagger (n)}\vec{D} 
  \Omega_{\rm es,edm}\m \Phi_0 \ra \bigg ) .
\end{equation}
Which can be rewritten in terms of CSF coefficients as
\begin{equation}
  d_a = 2\bigg [\la \Phi_0 \m +\sum_i\la \Phi_i \m 
  {\cal C}_i^{\rm (es)}\bigg ]\vec{D} \Omega_{\rm es,edm}\m \Phi_0 \ra  
  = 2\sum_j\bigg [\la \Phi_0 \m +\sum_{i\ne 0}\la \Phi_i \m 
  {\cal C}_i^{\rm (es)}\bigg ] \vec{D} {\cal C}_j^{\rm (es,edm)}\m 
  \overline{\Phi}_j \ra  .
 \label{eq5.1}
\end{equation}
Where the definitions of all the quantities are the same as defined in paper-I. 
Within the whole CSF space, the contribution from the first term is $5.4394\;39$
and the contribution from the second term is $-0.6417\;31$, which is just 
11.80\% of the first term. Five most important configurations in the second term
from the even-parity sub-space are $|6p*^2\r$,  $|6p^2\r$, $|6p\!*7p*\r$, 
$|5d*^2\r$ and $|6p7p\r$ and their contributions are $-0.8119\;50$, $0.2065\;15$,
$-0.1001\;58$, $0.0738\;55$ and $0.0599\;84$ respectively, where $d_a$ is in 
units of $C_T\sigma_N\times 10^{-12}ea_0$. All these are doubly excited 
CSFs and mixes with the ground state CSF significantly but the singly excited CSF
$|6s7s\r$ which is the third important CSF of $|\Psi_0\r$ does not contribute
strongly. In addition there is shift in the sequence of the important CSFs 
compared to the sequence of contribution to $|\Psi_0\r$, this is due
to difference of  dipole and \edm \ coupling strengths between different CSFs.

 Within the whole CSF space considered the value of $d_a$ is 
$4.4438\;58$. In absolute terms this is $0.4804\;56$ less compared to the 
lowest order result of $4.9243\;136$ calculated in an earlier section. A major
contribution to this difference is the many-body effects, this is because the
direct contributions from the CSFs added to the configuration space is small.
Which implies that the contribution from the many-body effects is just 10.81\%
of the total value and the change is negative. An added  advantage of 
the order by order approach is that the contribution to $d_a$ can be calculated
in terms of the order of residual coulomb interaction. Earlier while calculating
the ground state ASF energy it was shown how a truncation in the order of $\he$
perturbation can give an inaccurate value of $E_0$.


\subsection{Cluster Based Formulations }

 The CEPA-0 wave-operator equations of $\Omega_{es}$ and  $\Omega_{es,edm}$ 
are identical to the linearized coupled-cluster equations.  The atomic Yb EDM
$d_a$ calculated using CEPA-0 using the same sequence of CSFs as in 
Table \ref{table3} are given in Table \ref{table4} and the following can be 
inferred:
\begin{enumerate}
  \item The results from CEPA-0 do not differ significantly from the MBPT 
        results for the singly excited CSFs with core configurations 
        $|6s\rangle$ and $|4f^{13}6s^2\rangle$. This is because the non-linear 
        terms do not contribute since the configuration space is spanned by 
        singly excited CSFs. 
  \item A significant difference from the MBPT results is expected when 
        doubly excited CSFs are included in the CSF space. This is 
        observed when the CSFs having core configuration $|4f^{13}6s^1\rangle$ 
        are included. The change of $d_a$ calculated using MBPT by fixing the 
        number of odd parity CSFs to 17087 is $-0.5599\,84$, whereas CEPA-0
        reduces the change to $-0.1091\,50$.
  \item Within the whole CSF space considered, the value of $d_a$ calculated 
        using CEPA-0 is $5.9421\,36$, which is 25.21\%  larger than the
        MBPT result. The difference gives a rough estimate of the contribution
        from the non-linear terms in the cluster equation. It is a rough
        estimate as the MBPT calculation also includes the contribution from 
        the size-inconsistent terms, which are unphysical and are not included
        in the CEPA formalism.
\end{enumerate}
The CEPA-0 calculation does not include size-consistent non-linear terms of
$\Omega_{es}$ and  $\Omega_{es,edm}$. The results using the CEPA-2 equations are
tabulated in Table \ref{table5}. To make comparisons convenient the calculation 
is done with the same choice of CSF sequence as before. Comparing with the 
results from the calculations using MBPT and CEPA-0 give the following: 
\begin{enumerate}
  \item The  calculation within the CSF space of singly excited even parity CSFs
        with core configurations $|6s\rangle$ and $|4f^{13}6s\rangle$ and all the
        odd parity CSFs is identical with the result calculated using CEPA-0. 
        This implies that the contribution from the non-linear EPV terms with 
        $T_{\rm es}(1)$ from the even parity subspace is negligible. But, it is 
        different from the MBPT result of the same CSF space. And the 
        difference of $0.2067\,67$ is due to the EPO non-linear terms, which are 
        included in the MBPT.
  \item Consider the calculation  within the CSF space consisting of all the 
        even parity CSFs and the singly excited odd parity CSFs having core
        configurations $|6s\rangle$ and $|4f^{13}6s\rangle$. The CEPA-2 and 
        MBPT results differ by less than 2\%. But, the result from CEPA-0 
        differs from both by more than 12\%. This  shows that the contribution 
        from non-linear terms with $T_{\rm PTV}(1)$ is not negligible or the
        contribution from the size-consistent is significant, with the present
        results it is difficult to distinguish between the two. 
  \item The $d_a$ calculated within the whole CSF space considered is 
        $4.5065\,25$ and is larger than the MBPT result by 1.4\% and less than 
        the CEPA-0 result by 31.86\%. This implies that the contribution of 
        non-linear terms to $d_a$ is very important.
\end{enumerate}
The  effect of the EPO unlinked terms in $\Omega_{\rm es,edm}$ can be estimated 
by calculating $T_{\rm es,edm}$ with the EPO renormalization terms included. The
results  are given in Table \ref{table6}. The final value with the full 
CSF space is $4.2446\;91$, which is suppressed by 5.8\% compared to the CEPA-2 
result. This difference is due to the EPO contribution to the renormalization 
term in $\Omega_{\rm es,edm}$. Another property which can be compared to gain an
insight on the contribution of the size-inconsistent terms is the energy of the 
ground state $E_0$. The value of $E_0$ calculated using the wave-operator 
$\Omega_{\rm es}$ derived here has no contribution from the size inconsistent 
terms but it excludes some of the less important size-consistent terms. The 
difference in the value $E_0$ calculated  using the Bloch-equation and 
$\Omega_{\rm es}$ derived  from the cluster equation gives the contribution from 
the size inconsistent terms. Like in $d_a$ the approximation is that the 
contribution form the EPO size consistent terms non-linear in cluster amplitudes
is very small, then the difference in the result can in principle be accounted to 
the size inconsistent terms.


\section{Conclusion}

  Comparing the  results from different methods, it is clear that a 
size-consistent theory is preferable for high accuracy computation of atomic
EDM and the contribution from the non-linear terms in the cluster amplitude is
also important. With the present calculation the difference between the MBPT
and CEPA-2 result cannot be accounted distinctly to non-linear terms or the
size-inconsistent terms. A better comparison can be made after including 
all the non-linear terms in the cluster amplitudes. As the orbital space is 
made larger, the size of the CSF space grows very large. This puts a limitation 
during the calculation as the memory requirement increases, it is manageable if 
the calculation is done at the level of single particle--Moller-Plesset 
partitioning of the atomic Hamiltonian. At the single particle level the cluster
equation with the residual coulomb interaction Hamiltonian reduces to the 
method used used by Blundell \cite{blundell} and his collaborators. The 
inclusion of terms non-linear in cluster amplitudes is also relatively easy 
as compared to the EN-partitioning. To check the quality of the wave functions,
experimentally known quantities like hyperfine constants and excitation
energies needs to be calculated. These will be reported in our later papers.


\section{Acknowledgments}

  We thank our colleagues Rajat Choudury,  Holger Merlitz and P. Panda for
many useful discussions we had and the computer staff for making available to 
us the r10000 power challenge.


\begin{figure}[h]
  \makebox{
  \epsfig{height=3in,width=3.3in,figure=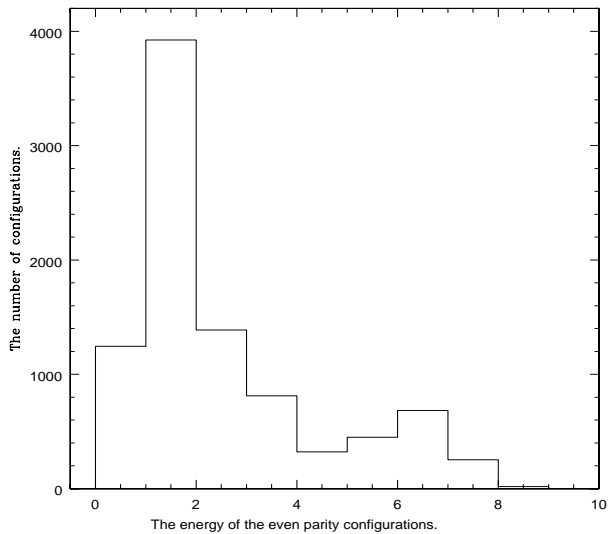} 
  \epsfig{height=3in,width=3.3in,figure=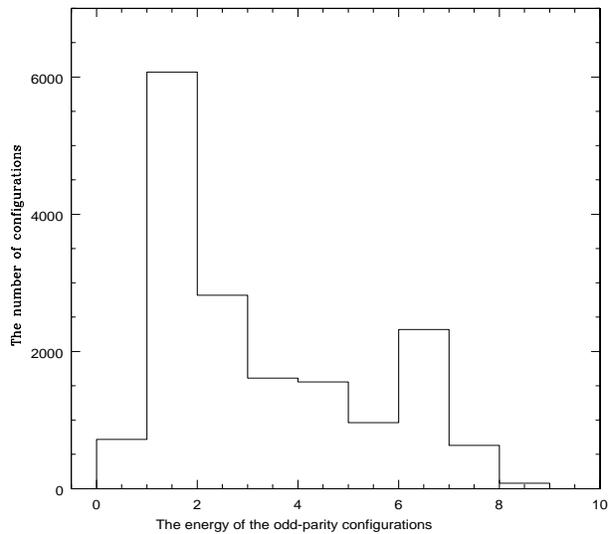}}
  \hspace*{1.5in}(a)\hspace*{3.2in}(b)
 \caption{Histogram of $|E|$ for the even parity CSFs.}
\end{figure}

\begin{figure}[h]
  \makebox{
  \epsfig{height=3in,width=3.3in,figure=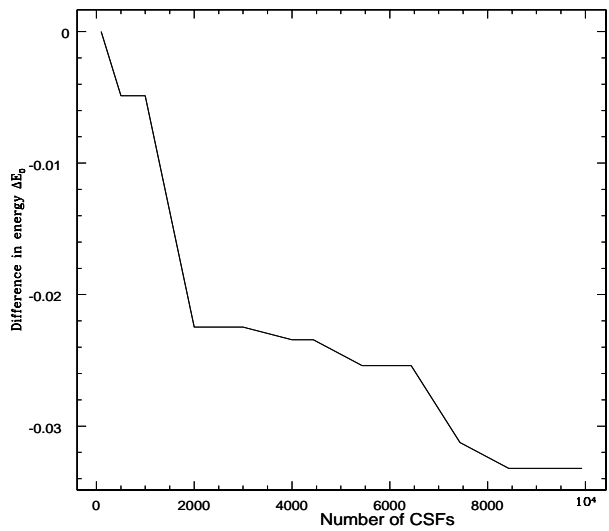}
  \epsfig{height=3in,width=3.3in,figure=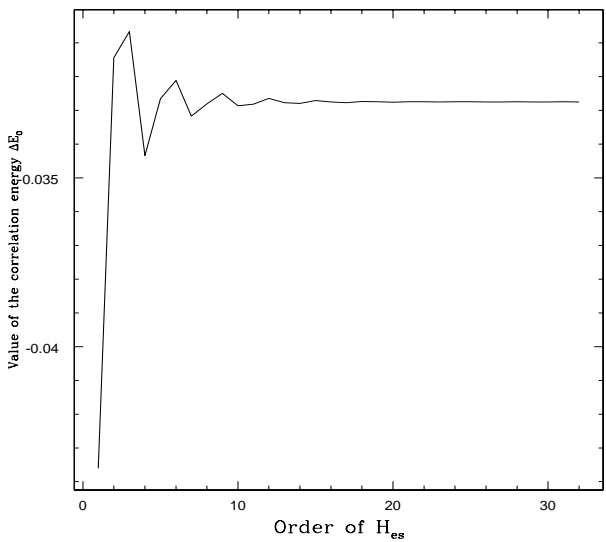}}
  \hspace*{1.5in}(a)\hspace*{3.2in}(b)
  \caption{The change in energy due to many-body effects introduced by the 
           configurations.}
  \label{fig7}
\end{figure}

\begin{figure}[h]
  \makebox{
  \epsfig{height=2.5in,width=3.3in,figure=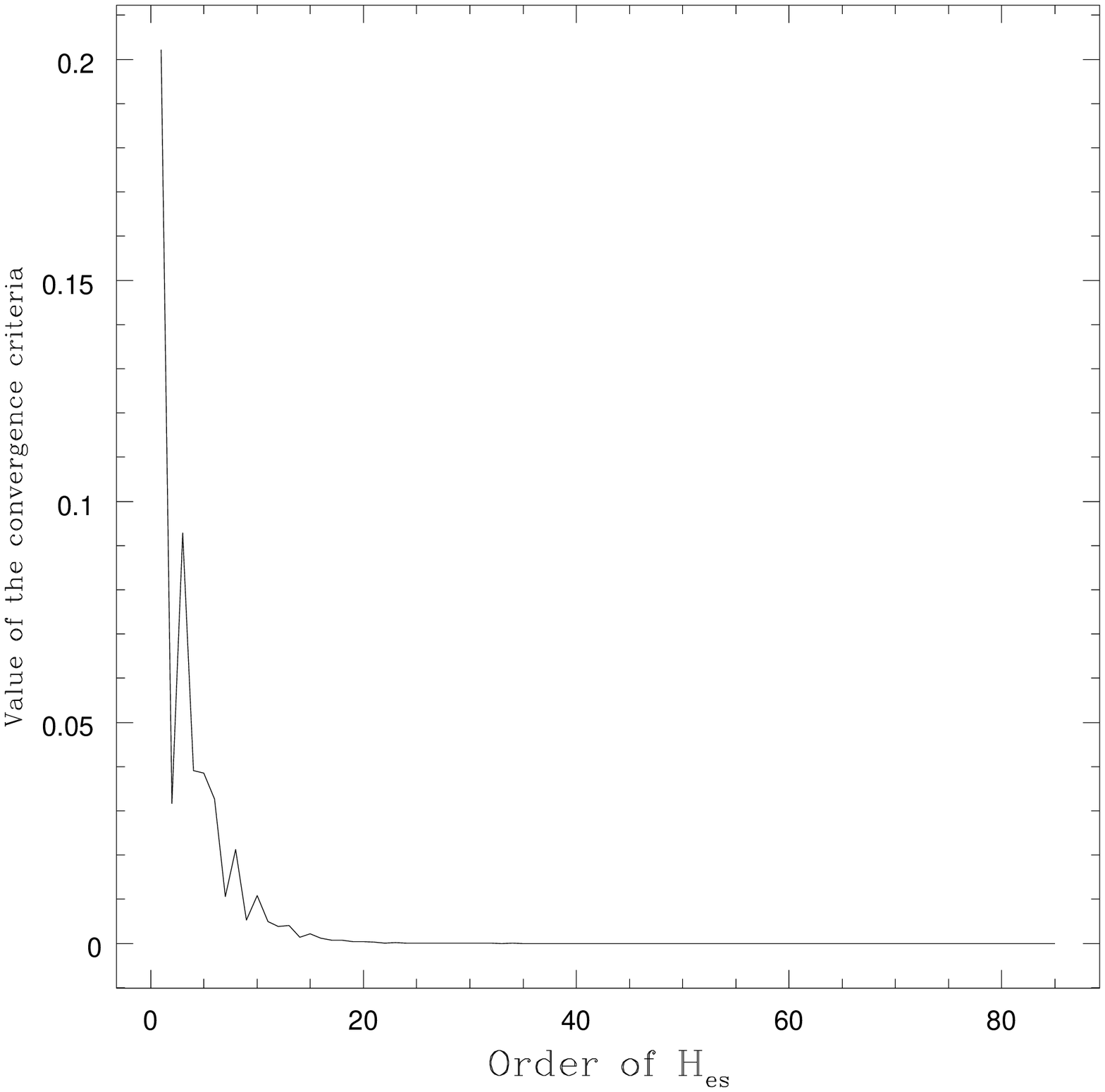} 
  \epsfig{height=2.5in,width=3.3in,figure=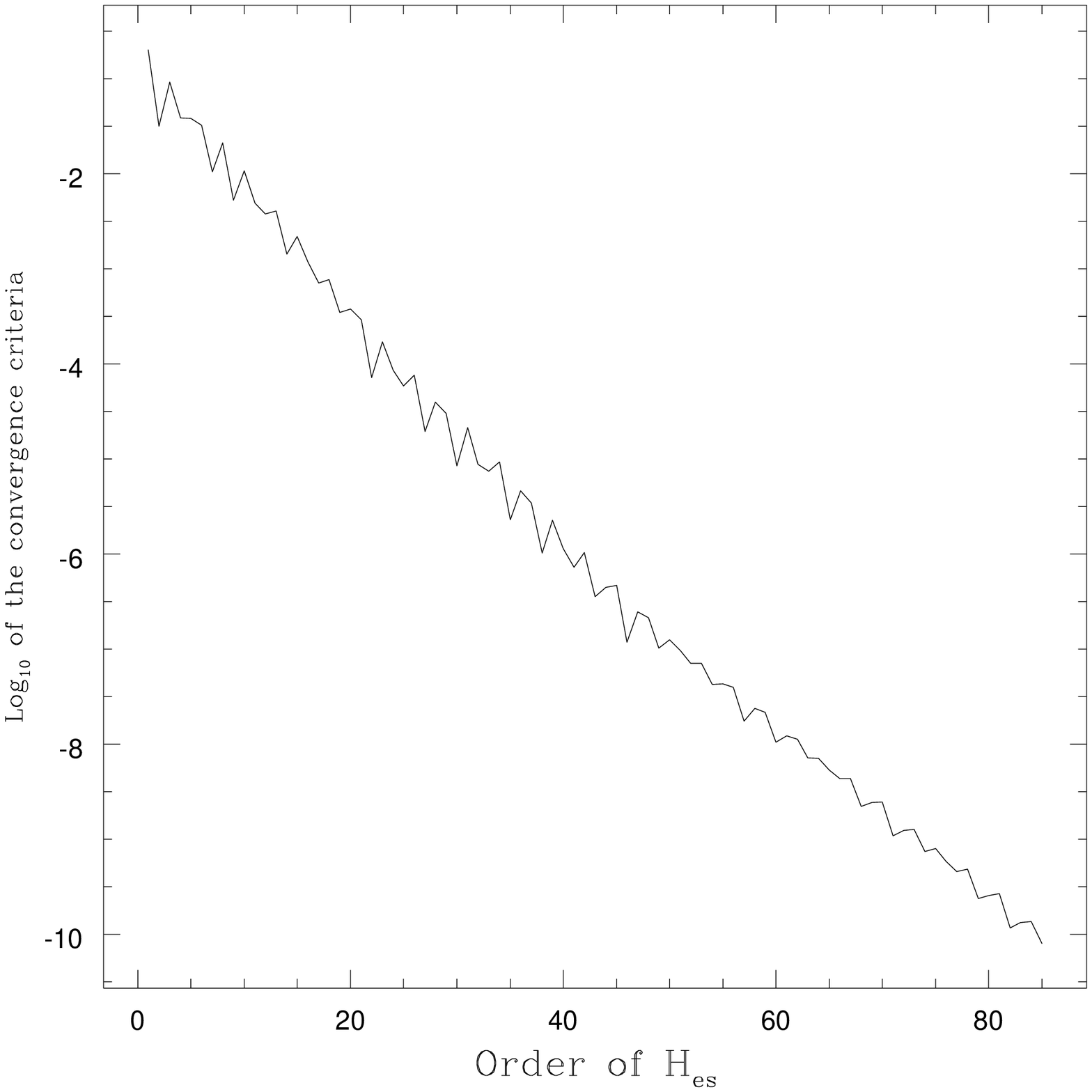}}
  \hspace*{1.5in}(a)\hspace*{3.2in}(b)
  \caption{The convergence of the wave-operators }
  \label{fig8}
\end{figure}

\begin{figure}[h]
  \makebox{
  \epsfig{height=2.5in,width=3.3in,figure=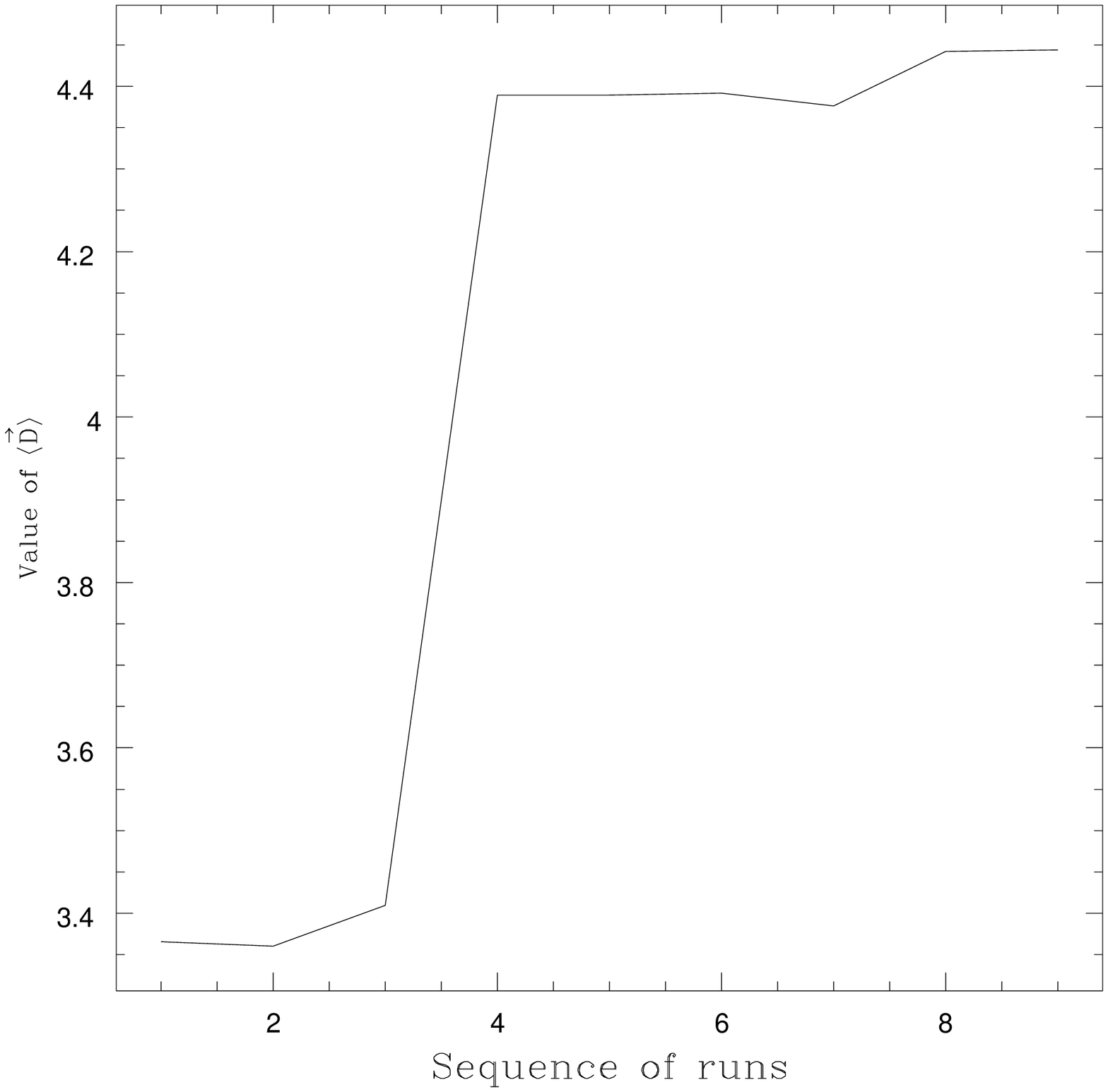} 
  \epsfig{height=2.5in,width=3.3in,figure=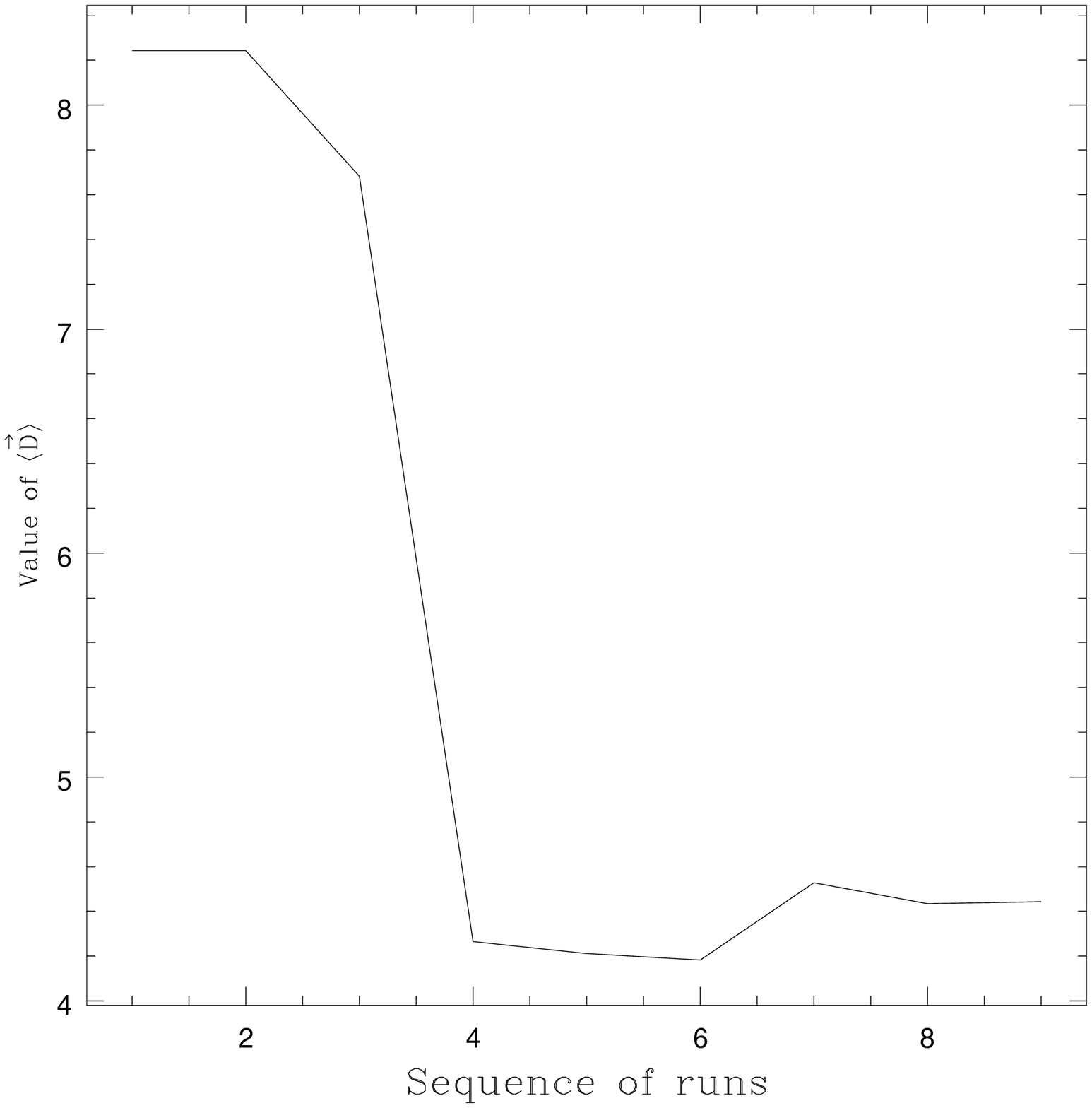}}
  \hspace*{1.5in}(a)\hspace*{3.2in}(b)
  \caption{ The value of $d_a$ }
  \label{fig9}
\end{figure}


\begin{table}[hbt]
 \caption{ The number of the CSFs with different occupied configurations.}
 \begin{tabular}{c|l|l|l|c|l|l|l} \hline \hline
 Sl.no & Occupied part           &\multicolumn{2}{c|}{Configurations}&
 Sl.no & Occupied part           &\multicolumn{2}{c}{Configurations} \\ 
                                                       \cline{3-4}\cline{7-8}
      &                     & Even     & Odd      &
      &                     & Even     & Odd      \\ \hline
  1   & $|6s\r$             & 6        & 12     & 
  2   & $|4f^{14}\r$        & 147      & 287     \\
  3   & $|4f^{13}6s^2\r$    & 12       & 18      &
  4   & $|4f^{13}6s^1\r$    & 1224     & 3618    \\
  5   & $|4f^{12}6s^2\r$    & 3045     & 7739    & 
  6   & $|5p^54f^{14}6s^2\r$& 12       & 30      \\
  7   & $|5p^54f^{14}6s^1\r$& 1044     & 1668    &  
  8   & $|5p^54f^{13}6s^2\r$& 3604     & 2394    \\ \hline
 \end{tabular} 
 \label{table1}
\end{table}

\begin{table}[h]
 \caption{ The energy of the ground state ASF with increasing CSF-space size.}
 \begin{tabular}{c|c|c|c|c|c}  \hline \hline 
  Sl. no &No of CSF& Energy          & Sl. no& No of CSF & Energy   \\ \hline
   1     & 100     &$-14067.6714\;79$&   2   & 500       &$-14067.6755\;69$ \\
   3     & 1000    &$-14067.6756\;60$&   3   & 2000      &$-14067.6942\;64$ \\
   5     & 3000    &$-14067.6942\;68$&   6   & 4000      &$-14067.6949\;91$ \\
   7     & 4435    &$-14067.6949\;91$&   8   & 5435      &$-14067.6974\;13$ \\
   9     & 6435    &$-14067.6974\;21$&  10   & 7435      &$-14067.7019\;28$ \\
  11     & 8435    &$-14067.7040\;97$&  12   & 9094      &$-14067.7040\;97$ \\
  13     & 9594    &$-14067.7042\;26$&  14   & 9930      &$-14067.7042\;26$ \\
 \hline
 \end{tabular} 
 \label{table2}
\end{table}

\begin{table}[h]
 \caption{ Values of $d_a$ for different number of even and odd parity
          configurations.}
 \begin{tabular}{c|c|c|c|c|c|c|c} \hline \hline
 Sl. no& \multicolumn{2}{c|}{No of CSFs} &$d_a$ & Sl. no & 
     \multicolumn{2} {c|}{No of CSFs} &$d_a$\\ \cline{2-3} \cline{6-7}
    & Even  & Odd   &             &    &Even    & Odd   &    \\ \hline
  1 & 9930  & 12    &$3.3654\;83$ & 10 & 7      & 17087 &$8.2430\;89$ \\
  2 & 9930  & 30    &$3.3605\;54$ & 11 & 19     & 17087 &$8.2430\;89$ \\
  3 & 9930  & 3648  &$3.4095\;46$ & 12 & 1243   & 17087 &$7.6831\;05$ \\
  4 & 9930  & 3846  &$4.3895\;21$ & 13 & 1390   & 17087 &$4.2645\;44$ \\
  5 & 9930  & 11676 &$4.3895\;43$ & 14 & 4435   & 17087 &$4.2118\;95$ \\
  6 & 9930  & 14070 &$4.3916\;11$ & 15 & 4448   & 17087 &$4.1837\;51$ \\
  7 & 9930  & 14100 &$4.3760\;30$ & 16 & 6713   & 17087 &$4.5282\;75$ \\
  8 & 9930  & 15768 &$4.4424\;65$ & 17 & 9094   & 17087 &$4.4333\;81$ \\
  9 & 9930  & 17087 &$4.4438\;58$ & 18 &        &       &             \\
 \hline
 \end{tabular} 
 \label{table3}
\end{table}

\begin{table}[h]
 \caption{ Value of $d_a$ computed using the CEPA-0 formalism.}
 \begin{tabular}{c|c|c|c|c|c|c|c} \hline \hline
 Sl. no & \multicolumn{2}{c|}{No of CSFs} &$d_a$ & Sl. no & 
     \multicolumn{2} {c|}{No of CSFs} &$d_a$\\ \cline{2-3} \cline{6-7}
     & Even  & Odd   &             &    & Even   & Odd   &    \\ \hline
   1 & 9930  & 12    &$3.8250\;23$ & 10 & 7      & 17087 &$8.4499\;64$ \\
   2 & 9930  & 30    &$3.8426\;96$ & 11 & 19     & 17087 &$8.4499\;64$ \\
   3 & 9930  & 3648  &$3.9252\;41$ & 12 & 1243   & 17087 &$8.3408\;14$ \\
   4 & 9930  & 3846  &$5.7783\;09$ & 13 & 1390   & 17087 &$5.5126\;26$ \\
   5 & 9930  & 11676 &$5.7783\;88$ & 14 & 4435   & 17087 &$5.5091\;16$ \\
   6 & 9930  & 14070 &$5.7820\;42$ & 15 & 4448   & 17087 &$5.3769\;11$ \\
   7 & 9930  & 14100 &$5.8370\;85$ & 16 & 6713   & 17087 &$5.6132\;71$ \\
   8 & 9930  & 15768 &$5.9437\;56$ & 17 & 9094   & 17087 &$5.9275\;26$ \\
   9 & 9930  & 17087 &$5.9421\;36$ &    &        &       &             \\
 \hline
 \end{tabular}
 \label{table4}
\end{table}

\begin{table}
 \caption{Table\ \thetable :\ Values of $d_a$ computed  with CEPA-2.}
 \begin{tabular}{c|c|c|c|c|c|c|c} \hline \hline
 Sl. no & \multicolumn{2}{c|}{No of CSFs} &$d_a$ & Sl. no & 
    \multicolumn{2} {c|}{No of CSFs} &$d_a$\\ \cline{2-3} \cline{6-7}
    & Even  & Odd   &             &    & Even   & Odd   &    \\ \hline
  1 & 9930  & 12    &$3.3130\;75$ & 10 & 7      & 17087 &$8.4499\;64$ \\
  2 & 9930  & 30    &$3.3133\;86$ & 11 & 19     & 17087 &$8.4499\;64$ \\
  3 & 9930  & 3648  &$3.3684\;39$ & 12 & 1243   & 17087 &$7.8528\;23$ \\
  4 & 9930  & 3846  &$4.4589\;69$ & 13 & 1390   & 17087 &$4.3791\;81$ \\
  5 & 9930  & 11676 &$4.4589\;33$ & 14 & 4435   & 17087 &$4.3758\;71$ \\
  6 & 9930  & 14070 &$4.4608\;15$ & 15 & 4448   & 17087 &$4.3558\;36$ \\
  7 & 9930  & 14100 &$4.4581\;78$ & 16 & 6713   & 17087 &$4.4547\;38$ \\
  8 & 9930  & 15768 &$4.5149\;07$ & 17 & 9094   & 17087 &$4.4981\;25$ \\
  9 & 9930  & 17087 &$4.5065\;25$ &    &        &       &             \\
 \hline
 \end{tabular} 
 \label{table5}
\end{table}

\begin{table}[h]
 \caption{Values of $d_a$ computed with the EPO components included in 
          the cluster amplitudes $\overline{\cal T}$.}
 \begin{tabular}{c|c|c|c|c|c|c|c} \hline \hline
 Sl. no & \multicolumn{2}{c|}{No of CSFs} &$d_a$ & Sl. no & 
     \multicolumn{2} {c|}{No of CSFs} &$d_a$\\ \cline{2-3} \cline{6-7}
     & Even  & Odd   &             &   & Even    & Odd   &    \\ \hline
   1 & 9930  & 12    &$3.1704\;91$ & 10 & 7      & 17087 &$8.4499\;64$ \\
   2 & 9930  & 30    &$3.1682\;67$ & 11 & 19     & 17087 &$8.4499\;64$ \\
   3 & 9930  & 3648  &$3.2196\;53$ & 12 & 1243   & 17087 &$7.8521\;84$ \\
   4 & 9930  & 3846  &$4.2130\;96$ & 13 & 1390   & 17087 &$4.3783\;45$ \\
   5 & 9930  & 11676 &$4.2130\;54$ & 14 & 4435   & 17087 &$4.3042\;71$ \\
   6 & 9930  & 14070 &$4.2147\;05$ & 15 & 4448   & 17087 &$4.2332\;39$ \\
   7 & 9930  & 14100 &$4.2039\;41$ & 16 & 6713   & 17087 &$4.2802\;30$ \\
   8 & 9930  & 15768 &$4.2539\;59$ & 17 & 9094   & 17087 &$4.2375\;23$ \\
   9 & 9930  & 17087 &$4.2446\;91$ & 18 &        &       &             \\
 \hline
 \end{tabular}
 \label{table6}
\end{table}

\end{document}